\pdfoutput=1

\documentclass[%
 reprint,
 amsmath,amssymb,
 aps,
twocolumn
]{revtex4-1}

\usepackage{hhline}
\usepackage{graphicx}
\usepackage{dcolumn}
\usepackage{bm}
\usepackage{siunitx}
\usepackage{caption}
\usepackage{subcaption}
\usepackage{nicefrac}

\usepackage{color}

\definecolor{darkgreen}{rgb}{0,0.5,0}

\begin{document}

\preprint{AIP/123-QED}


\title[]{Kinetics of 2D-constrained orbitally-shaken particles}

\author{Dhananjay Ipparthi}
\affiliation{IRIDIA, Universit\'{e} Libre de Bruxelles, 1050 Brussels, Belgium}%

\author{Tijmen A. G. Hageman}%
\affiliation{KIST Europe, Saarland University,  66123 Saarbr\"{u}cken, Germany}%
\affiliation{University of Twente, 7500 AE Enschede, The Netherlands}

\author{Nicolas Cambier}%
\affiliation{Universit\'{e} de Technologie de Compi\`{e}gne, 60200
  Compi\`{e}gne, France}%

\author{Metin Sitti}%
\affiliation{Physical Intelligence Department, Max Planck Institute
  for Intelligent Systems, and Max Planck ETH Center for Learning Systems, 70569 Stuttgart, Germany}%

\author{Marco Dorigo}%
\affiliation{IRIDIA, Universit\'{e} Libre de Bruxelles, 1050 Brussels, Belgium}%

\author{Leon Abelmann}%
\affiliation{KIST Europe, Saarland University, 66123 Saarbr\"{u}cken, Germany}%
\affiliation{University of Twente, 7500 AE Enschede, The Netherlands}

\author{Massimo Mastrangeli\textit{$^{\S}$}}%
\affiliation{Physical Intelligence Department, Max Planck Institute
  for Intelligent Systems, and Max Planck ETH Center for Learning
  Systems, 70569 Stuttgart, Germany}%

\footnotetext{\S~Current address: Electronic Components, Technology and Materials, Department of Microelectronics, Delft University of Technology, Delft, The Netherlands}


\begin{abstract}
We present an experimental study of the kinetics of orbitally-shaken sliding macroscopic particles confined to a two-dimensional space bounded by walls.
Discounting the forcing action of the external periodic actuation, the
particles undergo a qualitative transition from a ballistic to a
diffusive motion regime with time. 
Despite the deterministic input of kinetic energy provided by the shaker, the particles show translational velocities and diffusivity consistent with a confined random walk model.
Such experimental system may therefore represent a suitable macroscopic analogue to investigate aspects of molecular dynamics and self-assembly.
\end{abstract}

\keywords{self-assembly, macroscopic, kinetics, diffusion}
\maketitle

\section{\label{sec:introduction}Introduction}
Concepts pertaining to self-assembly can explain a variety of natural phenomena occurring across different scales, from molecular to macroscopic \cite{Klug, Whitesides:2002a,Whitesides:2002b}. 
An inherent difficulty in studying the self-assembly of molecular systems is posed by the very size of the self-assembling agents, generically defined here as \emph{particles}, and by the short duration of their interactions. 

As an alternative to fast spectroscopic
techniques~\cite{Calegari2014}, analogue microscopic
\cite{han2006brownian, kheifets2014observation, huang2011direct,
  loudet2004stokes, liu2016measuring, chakrabarty2014brownian, roy2016using} and macroscopic models \cite{Ilievski, hageman2018MacroEquivalence} of self-assembling systems can provide magnified if approximate representations of the interactions between particles, and of the time scales in which the interactions occur, amenable to easier investigations \cite{bian2016111}. 
Analog macroscopic models have proven useful to study at least two aspects of self-assembling systems: particles' kinetics \cite{Grzybowski} and population/concentration dynamics \cite{Hosokawa1994,Miyashita01042011,ipparthi2017yield}. 
Time evolution of particle populations has been theoretically studied using difference equations \cite{Gillespie_2007, Hosokawa1994,Miyashita01042011}.
The simulated results of these works show qualitative correspondence with experimental data obtained from  systems of macroscopic particles. 
Among available options~\cite{ojha2004,kudrolli2004,Kumar2015}, orbital shaking is a useful agitation method for macroscopic setups to study the dynamics and interactions of granular matter~\cite{Cademartiri2012,Tricard2013,Tricard2015,Hacohen2015,Bhalla2014,Grunwald2016}.
In particular, Scherer \emph{et al.} studied the swirling motion of rolling spherical particles~\cite{Scherer2000}.
The statistics of the motion that orbital shaking imparts to sliding polyhedral particles, bound to move by shear across a finite plane, was however not characterised to date.

In this paper we assess the motion of macroscopic, orbitally-shaken particles confined to a two-dimensional (2D) bounded space, and whether it may approximate the motion of molecules in a highly diluted 2D solution. 
The statistics of diffusing particles~\cite{gillespie2012} is described by a variety of random walk and Brownian motion models~\cite{einstein,smoluchowski,Langevin1908,bian2016111}. 
In our analogy with the microscopic realm, we adopt the confined Brownian motion model~\cite{bian2016111} as guiding framework to interpret our experimental data, though the mechanism underlying our particles' motion is admittedly different.

\section{Experimental setup}\label{sec:physical_system}

\begin{figure}[tbh]
	\centering
	\includegraphics[width=0.45\textwidth]{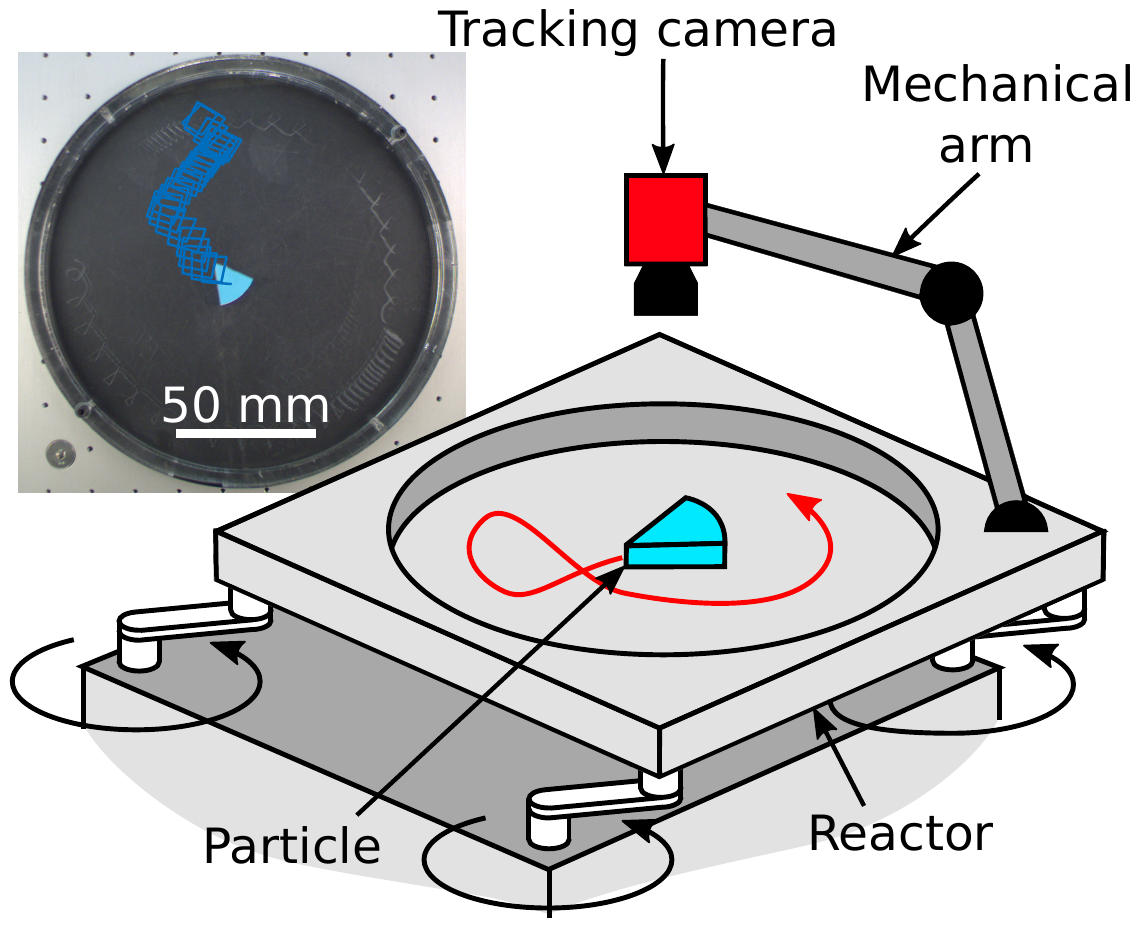}
	\caption{Schematic of the experimental setup used in this study.  
	A single particle and its trajectory, tracked through the overhanging camera solidal with the circular 2D reactor, are shown in the inset (see Fig.~\ref{fig:spectrum}b-c for the 3-particle case).}
	\label{fig:Setup}
\end{figure}

Our experimental setup was composed of a 2D circular \emph{reactor} confining 3D-printed particles, an orbital shaker (orbit diameter $d_\text{orb}= 2.5 \pm 0.1$ \si{cm})
imparting motion to the particles through the reactor, and an overhead camera (Basler acA1300-75gc, with a Basler C125-0618-5M F1.8 f6mm lens) for optical tracking (Fig.~\ref{fig:Setup}).
We used two studio diffused-light lamps inside a hood to uniformly illuminate the setup and avoid undesired shadows and reflections.
The particles were \SI{7}{\mm}-thick, geometrically equal sectors of a circle with radius of \SI{25}{\mm}.
The sectors spanned an angle of \SI{45}{\degree}, \emph{i.e.}, 8 of them could assemble to form a full circle~\cite{ipparthi2017yield}.
The particles' homogeneous colour and anisotropic shape were chosen to facilitate the tracking of their positions and angular orientations, respectively.
The particles were 3D printed by selective laser sintering using polyamide (PA 12).
The circular reactor was realised by cutting acrylic sheets.
The reactor had an inner diameter of \SI{25}{cm} and mildly rough interior surface, as we found surface roughness to reduce sliding friction and improve the mobility of the particles. 
The reactor was fixed onto the shaker using screws, and care was taken to ensure the reactor surface was levelled horizontal under stationary conditions. 
The camera was made solidal to the non-inertial frame of the shaker through a mechanical arm to avoid the need for shaker motion subtraction prior to image analysis.

The experiments were carried out under two conditions, whereby the motion of respectively one and three particles were tracked. 
Each experiment was performed by placing the particle(s) into the closed reactor, starting the video frame capture at \SI{20}{fps}, and then running the shaker at a frequency $f= 5.00 \pm 0.02$ Hz (\emph{i.e.}, 300 rpm).
Each experiment was run for \SI{8}{\minute} \SI{20}{\second} to capture \SI{10000}{frames}.  
Image processing of each acquired frame involved background subtraction, low-pass filtering and linear discriminant analysis to isolate the colour blobs corresponding to the particles based on their RGB value. 
Morphological operations were used to clean the blobs of artefacts. 
Within the shaker frame, the position $\textbf{x}$ of each particle was assumed to be the centre of mass of the corresponding blob. 
The orientation $\theta$ of each particle in the plane of the reactor was obtained by fitting lines onto the straight edges of the particles and evaluating the subtended angle in the shaker frame of reference.  

\section{Results}\label{sec:experiments_results}
We analysed three aspects of the particle kinetics: (1) velocity distribution, (2) translational diffusion and (3) rotational diffusion.  
Diffusion is the motion of particles caused by thermal energy~\cite{cooksy2014}.  
In the random walk model of Brownian motion, the translational velocities of particles with $k$ translational degrees of freedom are $\chi_k$-distributed~\cite{berg1993}.
The velocities of particles moving by Brownian motion in three dimensions follow a $\chi_3$ distribution, \emph{i.e.}, a Maxwell-Boltzmann distribution~\cite{berg1993}. 
Since in our case we constrained the particles to move in two dimensions ($k=2$), we might expect the 2D translational velocities and their 1D projections to respectively follow a $\chi_2$ distribution --- \emph{i.e.}, a Rayleigh distribution:
 \begin{equation}
	\label{eq:Ray}
	f(v,\sigma)=\frac{v}{\sigma^2}e^{-v^2/(2\sigma^2)}
 \end{equation} 
with velocity $v$ and corresponding mode $\sigma$ --- and a $\chi_1$ distribution, \emph{i.e.}, a Gaussian distribution, if a diffusion analogy holds for our system.

\begin{figure}[tbh]
	\centering
	\includegraphics[width=0.45\textwidth]{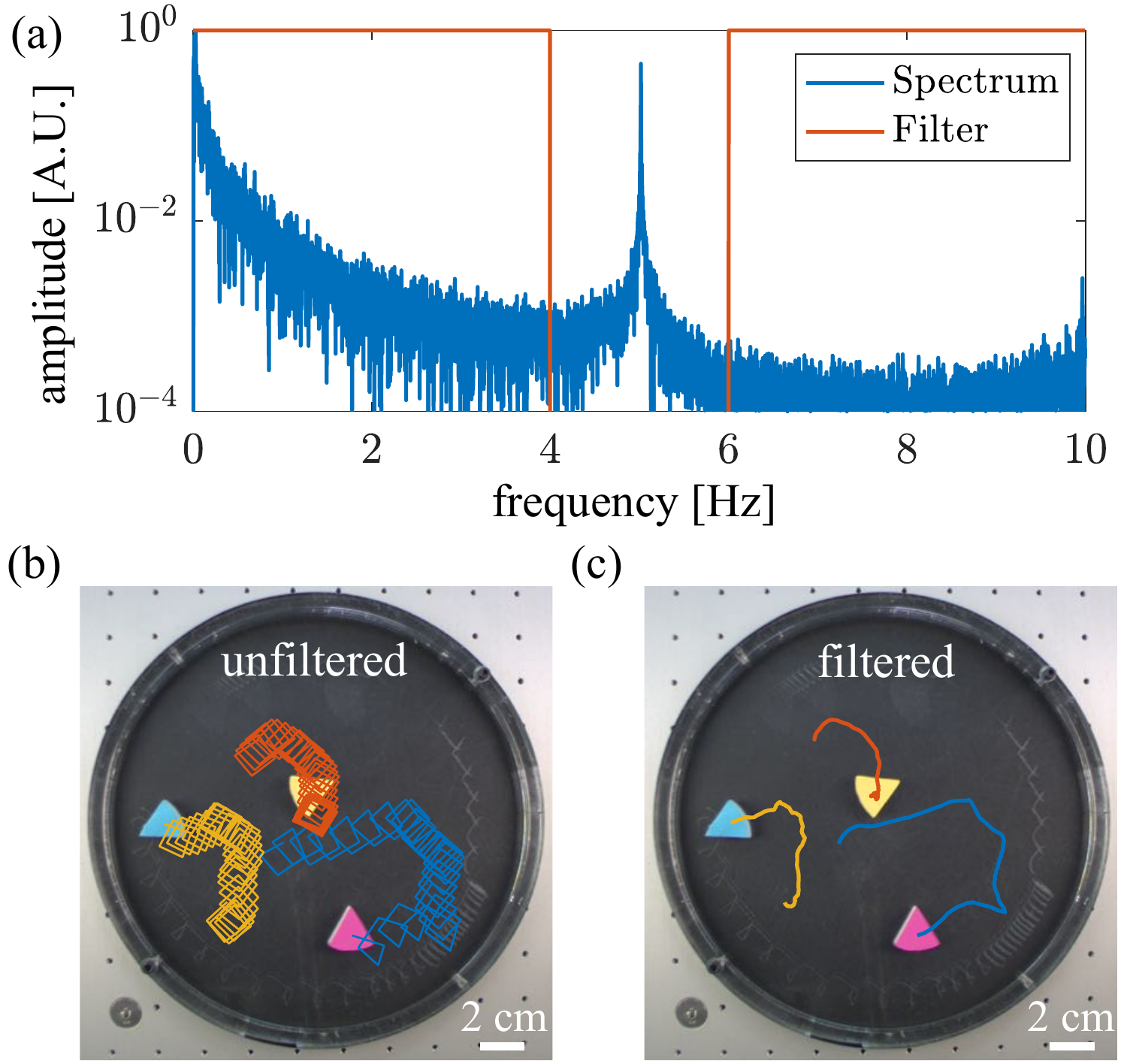}
	\caption{(a) Typical spectrum of the \emph{x}-coordinate of a particle,
	normalised to a maximum amplitude of 1, along with the band-stop
	filter excluding the actuation frequency of the orbital shaker.
	(b) Unfiltered and (c) filtered trajectories of 3 particles
        recorded over a period of \SI{5}{\s} (video available in Supplemental Material).
	}
	\label{fig:spectrum}
\end{figure}

In the kinetic analysis, we first considered the constant orbital component of the particle motion, expressed in the form of repetitive, short-range circular trajectories superposed to the piece-wise linear, long-range displacements of the particles (Fig.~\ref{fig:Setup}-inset and Fig.~\ref{fig:spectrum}b). 
Such orbital motion is due to the global actuation forced by the shaker and affects all particles equally and synchronously.
Therefore, while causing the motion of the particles by sliding friction, the orbital motion does not primarily contribute to particle interactions in a sparse particle system.
Particle interactions are mainly due to relative motion differences, as induced by, \emph{e.g.},~mutual collisions, local spatial inhomogeneities and boundary effects.
We discounted for the constant orbital component in the analysis of particle trajectories to focus on relative particle motion. 
Therefore, the actuation frequency was filtered out of the data by an FFT transform  of the $x$- and $y$- components of the trajectories into the frequency domain, after which the frequency components around the actuation frequency were eliminated by a rectangular band-stop filter (Fig.~\ref{fig:spectrum}a).
An inverse-FFT then transformed the results back in the time-space domain (Fig.~\ref{fig:spectrum}c).
The filter, with centre frequency of \SI{5}{\Hz} and width of about \SI{2}{\Hz} (Fig.~\ref{fig:spectrum}a), only suppressed the actuation frequency, and did not alter the phase behaviour of the particle motion. Analysis shows that the chosen filter width does not significantly affect the kinetic analysis (see discussion in Section \ref{transdiff}).
After filtering, the motion of the particles was mainly characterised
by low frequencies, whose normalised amplitude decayed by a factor
\num{100} at \SI{1}{\Hz} as compared to the unfiltered data. 

\begin{figure}[tbh]
	\centering
	\includegraphics[width=0.4\textwidth]{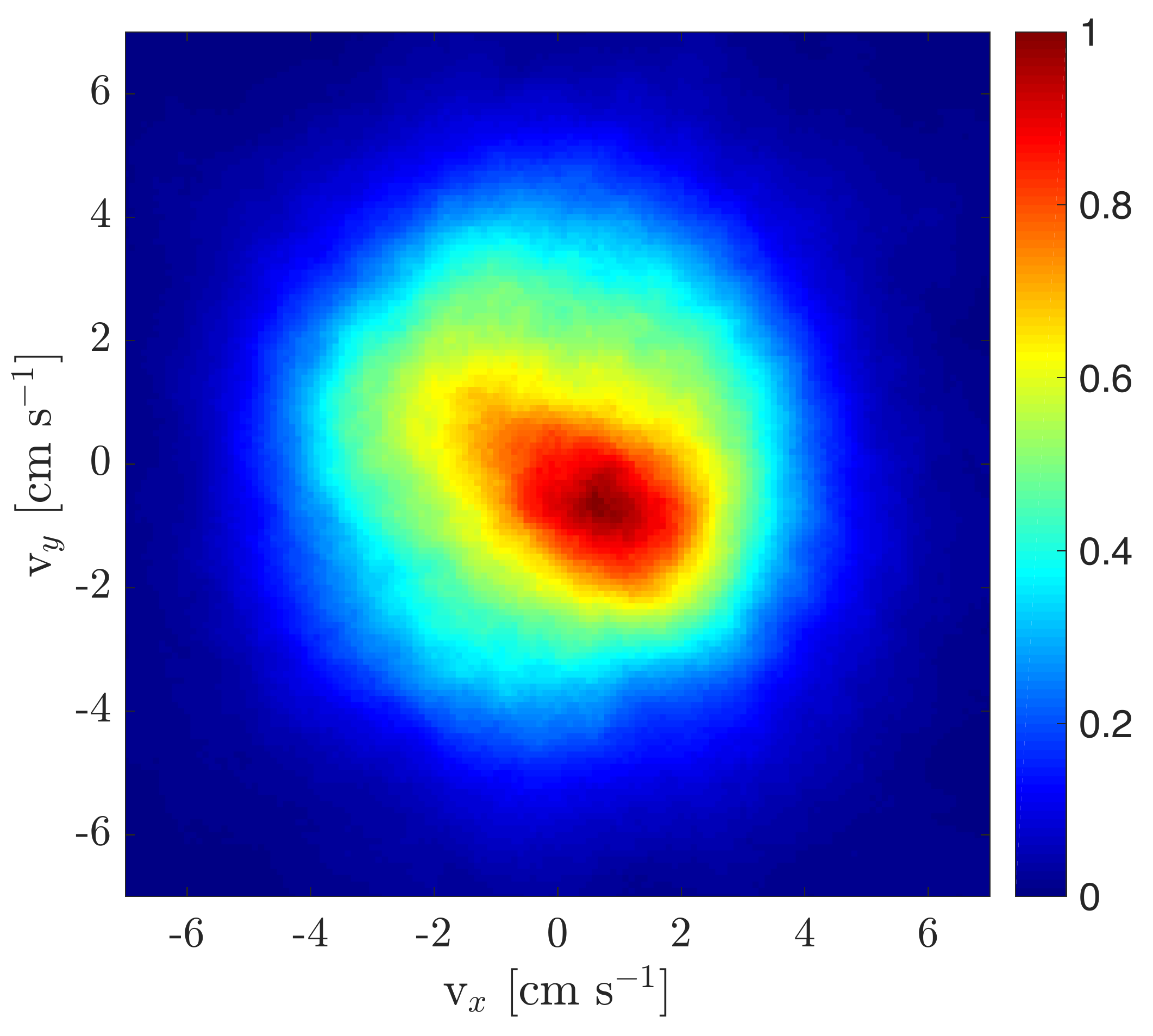}
	\caption{Normalised kernel density approximation of the \emph{x}- and \emph{y}-components of the
	measured filtered velocity of a single particle in the reactor.}
	\label{fig:v_heatMap_bin1_part2}
\end{figure}

The calculated values of cross-correlation between (filtered) \emph{x}- and \emph{y}-velocities of the 3-particle experiment are $-0.16, -0.18$ and $-0.04$ respectively, indicating the existence of a weak to very weak anti-cross-correlation.
Fig.~\ref{fig:v_heatMap_bin1_part2} displays the correlation
visually for the case of a single particle. It was generated by performing a kernel density
approximation of all data  points corresponding to the \emph{x}- and
\emph{y}-components of the velocity and normalising to $[0,1]$. 
A truly uncorrelated system is rotationally symmetric, while our
system shows a bias along the diagonal resulting from weak correlation.

\subsection{Velocity distribution}

\begin{figure}[tbh]
	\centering
	\includegraphics[width=0.45\textwidth]{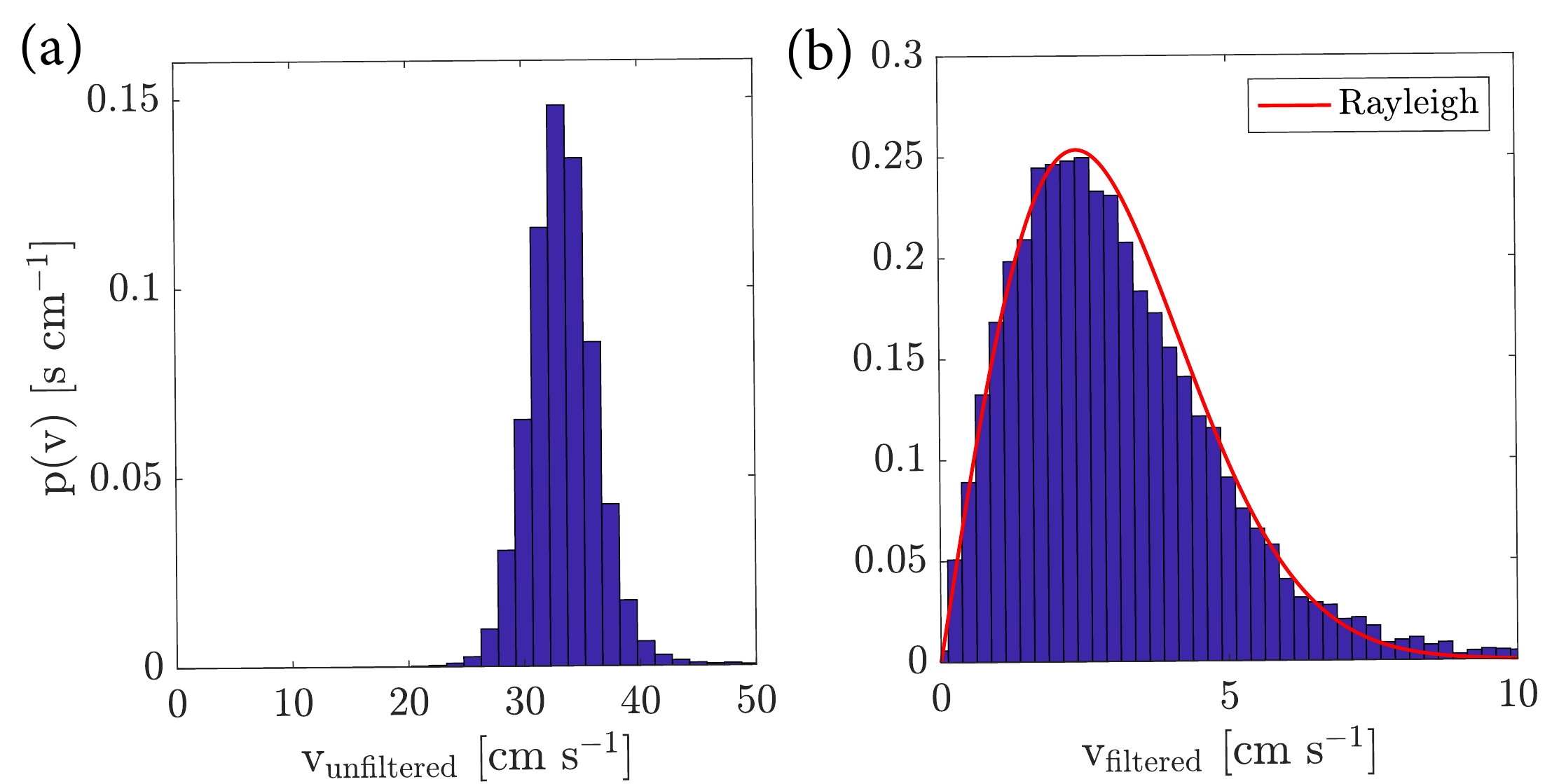}
	\caption{Typical 2D velocity distribution for (a) an unfiltered and (b) a filtered
          trajectory of a single particle tracked in the reactor.}
	\label{fig:Velocity2D}
\end{figure}

Fig.~\ref{fig:Velocity2D}a shows the typical unfiltered 2D velocity distribution of a single orbitally-shaken particle tracked in the reactor, with mean of
$33.42 \pm 0.03$ \si{\centi\meter\per\second} and a standard deviation of \SI{2.9}{\centi\meter\per\second}.
The mean particle velocity is expectedly close to the maximal orbital
speed afforded by the shaker, \emph{i.e.}, $v_\text{orb}\approx \pi f
d_\text{orb} = 39 \pm 2$ \si{\centi\meter\per\second}.

The filtered 2D particle velocity distribution appears to be Rayleigh-like (Fig.~\ref{fig:Velocity2D}b). 
A fitting routine for the Rayleigh distribution (Eq.~\ref{eq:Ray}) was used which minimises the maximum distance $D_\text{max}$ between the cumulative distribution (CDF) of the velocity measurements and the cumulative Rayleigh distribution, yielding the fitting parameter $\sigma$ and its uncertainty $\epsilon_\sigma$ (see Table~\ref{tab:fitting_values}). 
The CDFs are shown in Fig.~A1~\footnote{Figs.~A1, B1, C1 and D1 are available in the Appendix.}.
We tested whether our distribution is indeed Rayleigh using the Kolmogorov-Smirnoff (K-S) test, obtaining a significance level Q ~\cite{Press1992}.

The results, shown in Table~\ref{tab:fitting_values}, indicate that this is not the case (see very low Q values).
However, the maximum deviation $D_\text{max}$ of the measured cumulative velocity distribution from the cumulative Rayleigh distribution (Eq.~\ref{eq:Ray}) is $2.6\%$.
Since we analysed as many as \num{9600} individual tracks, the statistics is such that we can resolve a small difference between the measured and the theoretical distributions.
Such small deviation is quite remarkable, given that the driving force for the random walk of our macroscopic particles is fundamentally different from that of Brownian particles on the microscale. 
The deviation could be caused by the presence of hard reactor boundaries.  
Upon collisions, the reactor edges impulsively transmit kinetic energy to the particles and raise their velocities. 
Closer examination of Fig.~\ref{fig:Velocity2D}b indeed suggests that the observed velocities in the tail of the experimental distribution are consistently higher than in the fitted Rayleigh distribution, as also evident in Fig.~A1. 
Fig.~B1-a shows the unfiltered velocities plotted against the distance from the centre of the reactor, and Fig.~B1-b evidences that higher velocities were observed along the edges of the reactor.  
We cannot, however, exclude that local variations in roughness and flatness of the reactor surface play a role as well.

\begin{table}
	\caption{Fitting of the velocity distributions. Values for
          velocity distribution parameter $\sigma$ with uncertainty
          $\epsilon_\sigma$, and for significance level $Q$
          corresponding to the error measure $D_\text{max}$ obtained
          from the K-S test.}
	\label{tab:fitting_values}
	\begin{tabular}{cccccc}
		Samples & Dimensionality & $\sigma$ & $\epsilon_\sigma$ & $D_\text{max}$ & Q \\ 
			• & • & [\si{\centi\meter\per\second}] &[\si{\centi\meter\per\second}] & [\si{\percent}] & • \\
			\hhline{======}
			All & 2D & 2.45 & 0.02 & 2.6 & 4E-6 \\ 
			• & 1D$_x$ & 2.48 & 0.03 & 1.3 & 9E-2 \\ 
			• & 1D$_y$ & 2.34 & 0.04 & 2.1 & 4E-4 \\  
	\end{tabular} 
	\vspace{-20pt}
\end{table}

\begin{figure}[tbh]
	\centering
	\includegraphics[width=0.45\textwidth]{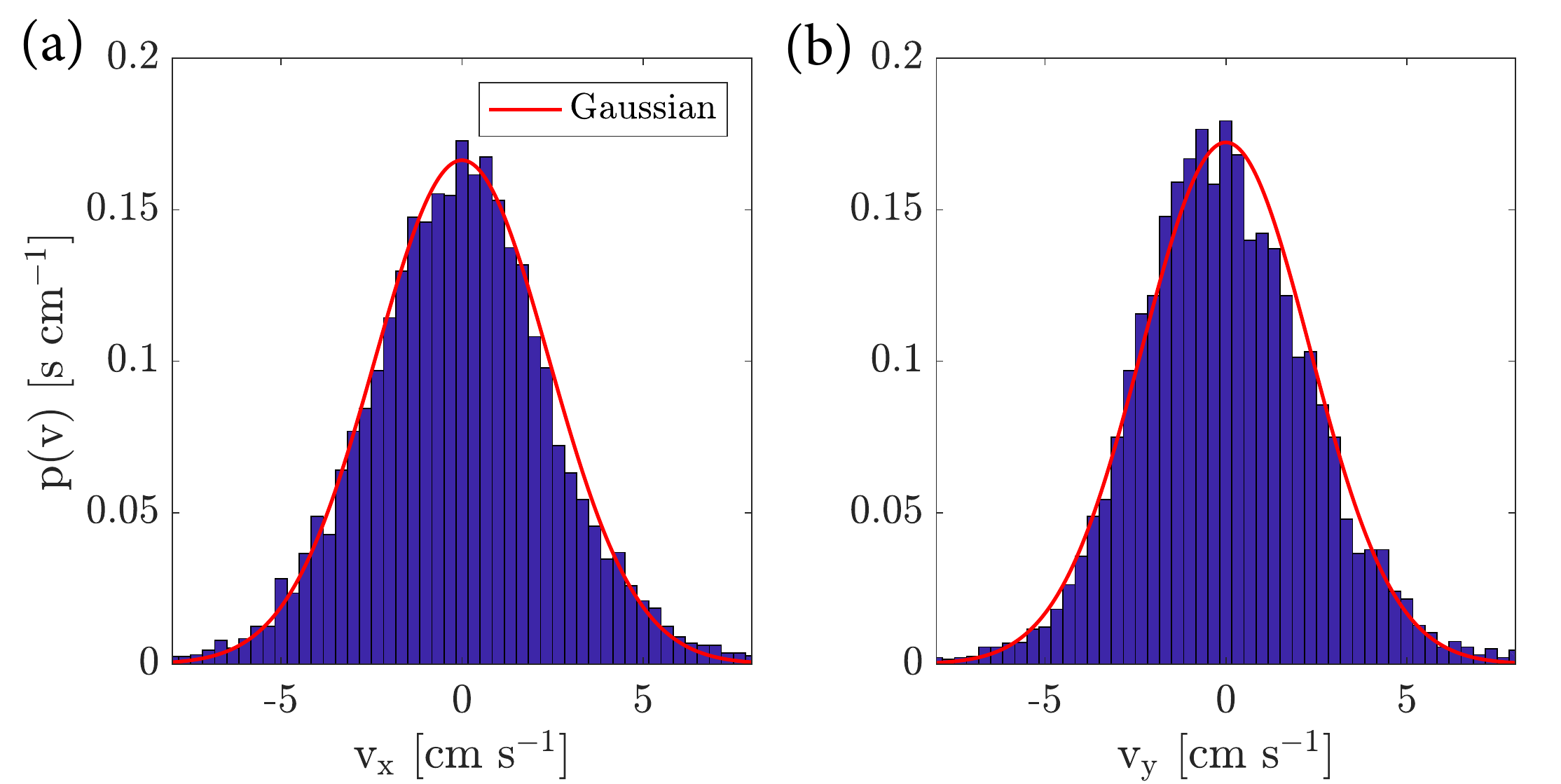}
	\caption{Distribution of the $x$- (a) and $y$-component (b) of the velocity of the single tracked particle with 2D velocity distribution shown in Fig.~\ref{fig:Velocity2D}-right.}
	\label{fig:Velocity2}
\end{figure}

Following our working framework, we might furthermore expect the distributions of the 2D particle velocity projected over orthogonal, one-dimensional axes to be Gaussian (\emph{i.e.}, $\chi_1$) distributions, and with similar velocity mode.
The 1D-projected velocity distributions for a single particle in the reactor are shown in Fig.~\ref{fig:Velocity2}, together with the Gaussian fit obtained using the same fitting procedure described earlier.
The results of the K-S test are also presented in Table \ref{tab:fitting_values}. 
The fitting parameter $\sigma$ for the 1D and 2D velocity distributions are very similar, which is in
agreement with the expected $\chi_k$-distribution.

The velocity distributions and corresponding results for the K-S test, with parameter $\chi_k$, for the 3-particle experiment, presented in Fig.~C1, show a more pronounced deviation from an ideal Brownian behaviour, possibly reflecting the effect of inter-particle collisions.

\subsection{Translational diffusion}\label{transdiff}

To study the diffusion of the particles in the reactor, we considered their complete (filtered) trajectory, divided it into equal-length sub-trajectories, and computed averages of the square displacement for every sampled time. 
The displacement was defined as the Euclidean distance from the starting point of the trajectory. 

For a two-dimensional diffusional motion, we might expect the following relation: 

\begin{equation}
	\left< X^2 \right> = 4 D t
\end{equation}

\begin{figure*}
    \centering
    \begin{subfigure}[b]{0.24\textwidth}
        \includegraphics[width=\textwidth]{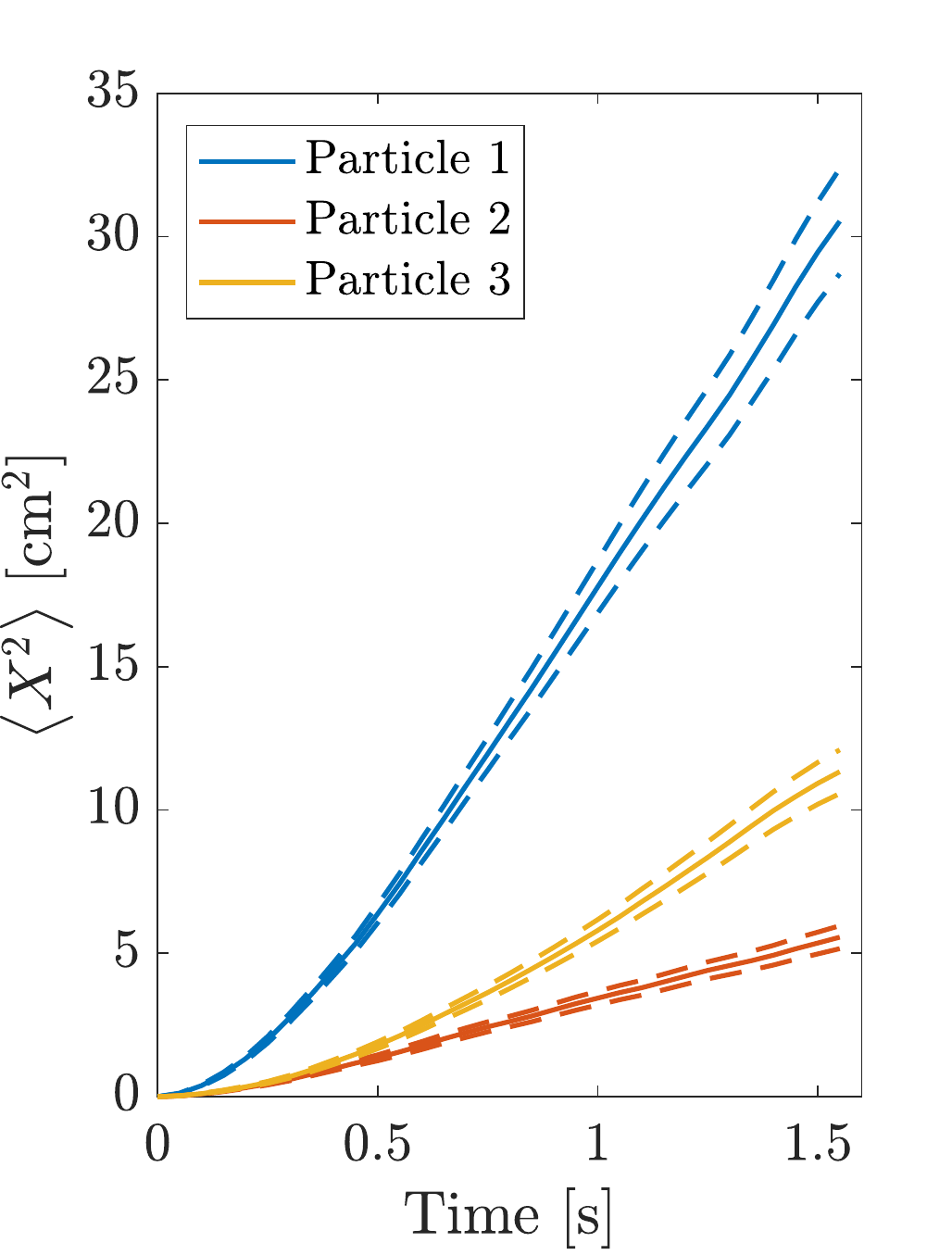}
        \caption{}
        \label{fig:Diffusion_3spheres_short}
    \end{subfigure}
    \begin{subfigure}[b]{0.24\textwidth}
        \includegraphics[width=\textwidth]{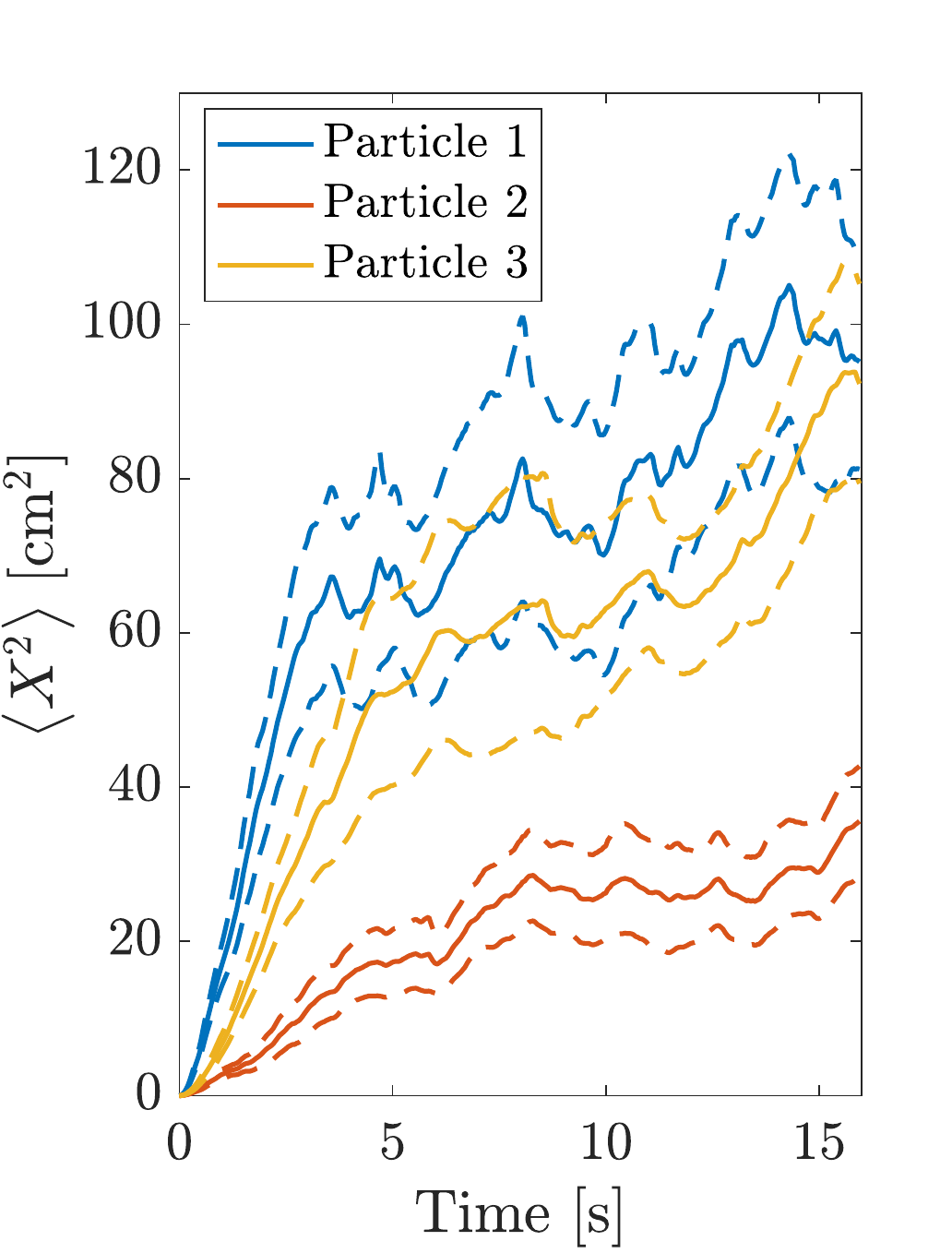}
        \caption{}
        \label{fig:Diffusion_3spheres_long}
    \end{subfigure}
    \begin{subfigure}[b]{0.24\textwidth}
        \includegraphics[width=\textwidth]{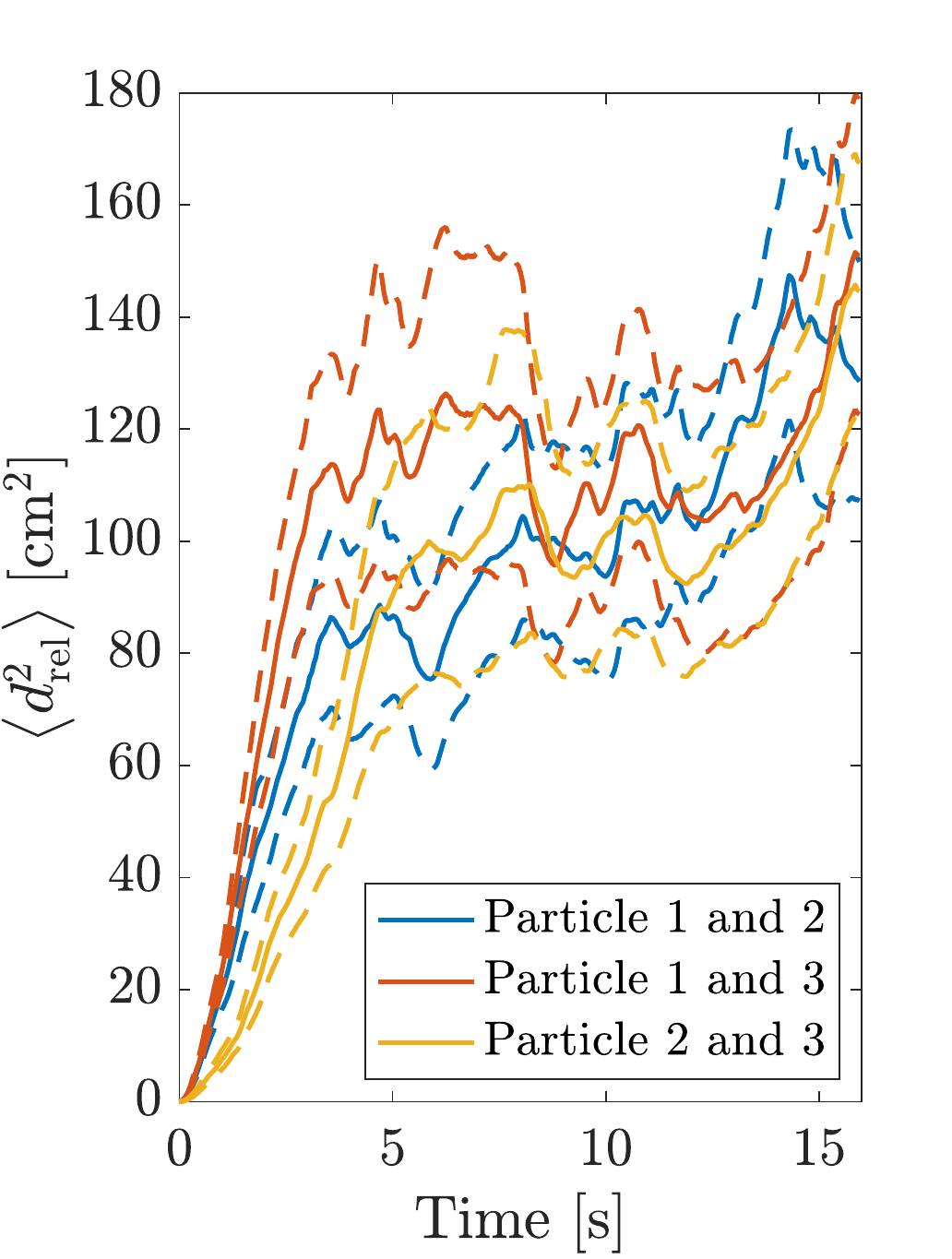}
        \caption{}
        \label{fig:Diffusion_3spheres_long_relative.pdf}
    \end{subfigure}
    \begin{subfigure}[b]{0.24\textwidth}
        \includegraphics[width=\textwidth]{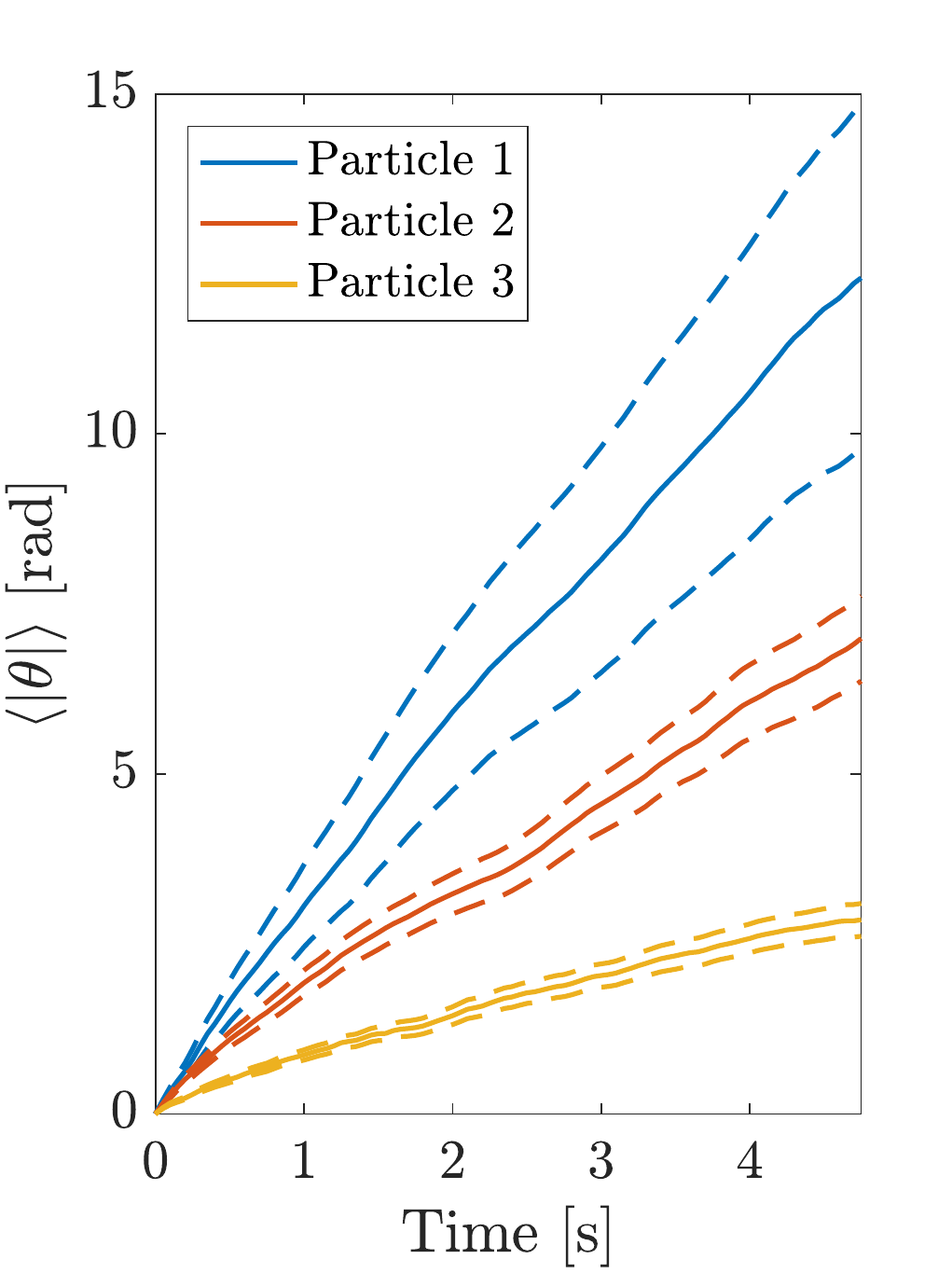}
        \caption{}
        \label{fig:Diffusion_3spheres_normal_angle_linear}
    \end{subfigure}
    \caption{Kinetics of three particles in the reactor. (a) Mean square displacement as a function of time ($t \leq $\SI{1.5}{\second}). The profile is indicative of the ballistic regime for $t \leq$ \SI{0.5}{\second}. (b) Mean square displacement as a function of time ($t \leq $\SI{15}{\second}). (c) Mean square relative displacement as function of time ($t \leq $\SI{15}{\second}). (d) Mean absolute angular displacement as a function of time ($t \leq $\SI{5}{\second}). In all plots, dashed lines indicate the standard error around the mean value.}
\end{figure*}

\noindent where $\left< X^2 \right>$ is the mean square displacement (MSD), $D$ the diffusion coefficient and $t$ the time \cite{einstein, perrin, smoluchowski}. 
Choosing more trajectories smoothens the data and decreases the standard error in estimating the true $\left< X^2 \right>$, but also decreases the observation time. 

For the 3-particle case, it can be seen from Fig.~\ref{fig:Diffusion_3spheres_short} that the particle motion starts off with a ballistic regime, characterised by a quadratic curve for $t<$ \SI{0.5}{\second}, before entering a linear regime. 
The diffusion coefficient, calculated by determining the slope of the linear regime using a $\chi$-square fitting method, varies from approximately
\SIrange{0.5}{5}{\square\centi\meter\per\second}.
The ensuing transition from the linear to a weak saturation regime is in accordance with the confined random walk principle \cite{perrin, smoluchowski}.  
The curves enter a saturation regime after roughly \SI{4 }{\second} (Fig.~\ref{fig:Diffusion_3spheres_long}). 

\begin{figure*}
	\centering
	\includegraphics[width=0.75\textwidth]{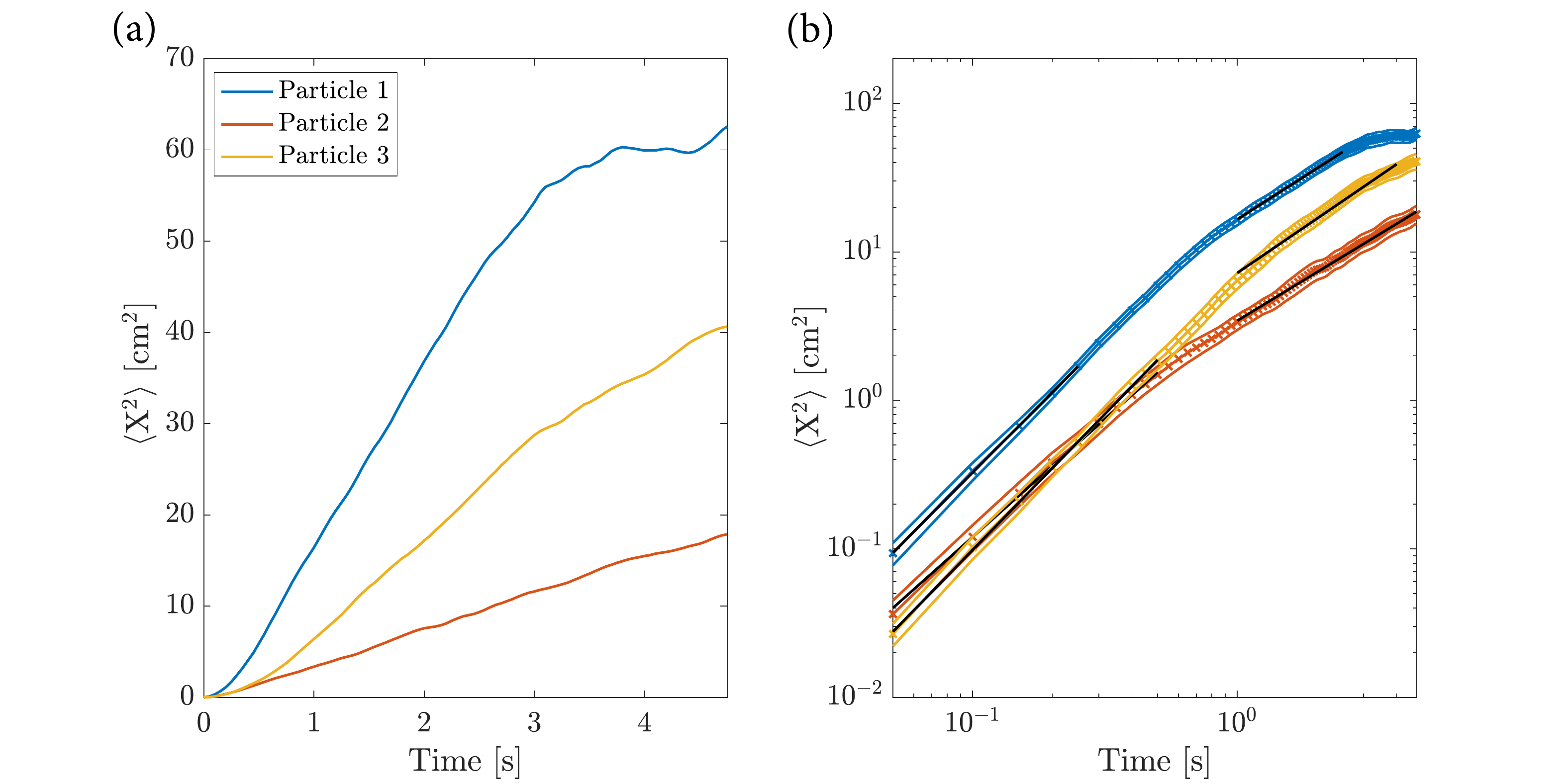}
	\caption{Mean square displacement on (a) linear and (b) double
          logarithmic plot, for the same time window of
          \SI{5}{\second}. In the double logarithmic plot the linear regions are interpreted as ballistic
          (short time scales) and diffusive (long time scale)
          regimes. The fitted slopes are tabulated in
          Table~\ref{table:approx_fitting_param_vals}. The filter width
          for both plots is \SI{2}{\Hz}.}
	\label{fig:X2_loglog}
\end{figure*}

To clarify the ballistic and linear regimes and emphasise the short
time scales, we plotted the MSD data both on a linear 
(Fig.~\ref{fig:X2_loglog}a) and on a double
logarithmic scale (Fig.~\ref{fig:X2_loglog}b). 
On a double logarithmic scale, a purely ballistic motion
would appear as a straight line with a slope of two, whereas pure unbounded diffusion would have a slope of one. 
In Fig.~\ref{fig:X2_loglog}b, the two regimes for a filter width of \SI{2}{Hz} are indicated by straight lines, and the slopes of the linear fits are listed in
Table~\ref{table:approx_fitting_param_vals}. 
Indeed, the slope for the linear fits at shorter time scales tend to be closer to two, while closer to one for the longer timescales.

\begin{figure*}
	\centering
	\includegraphics[width=0.75\textwidth]{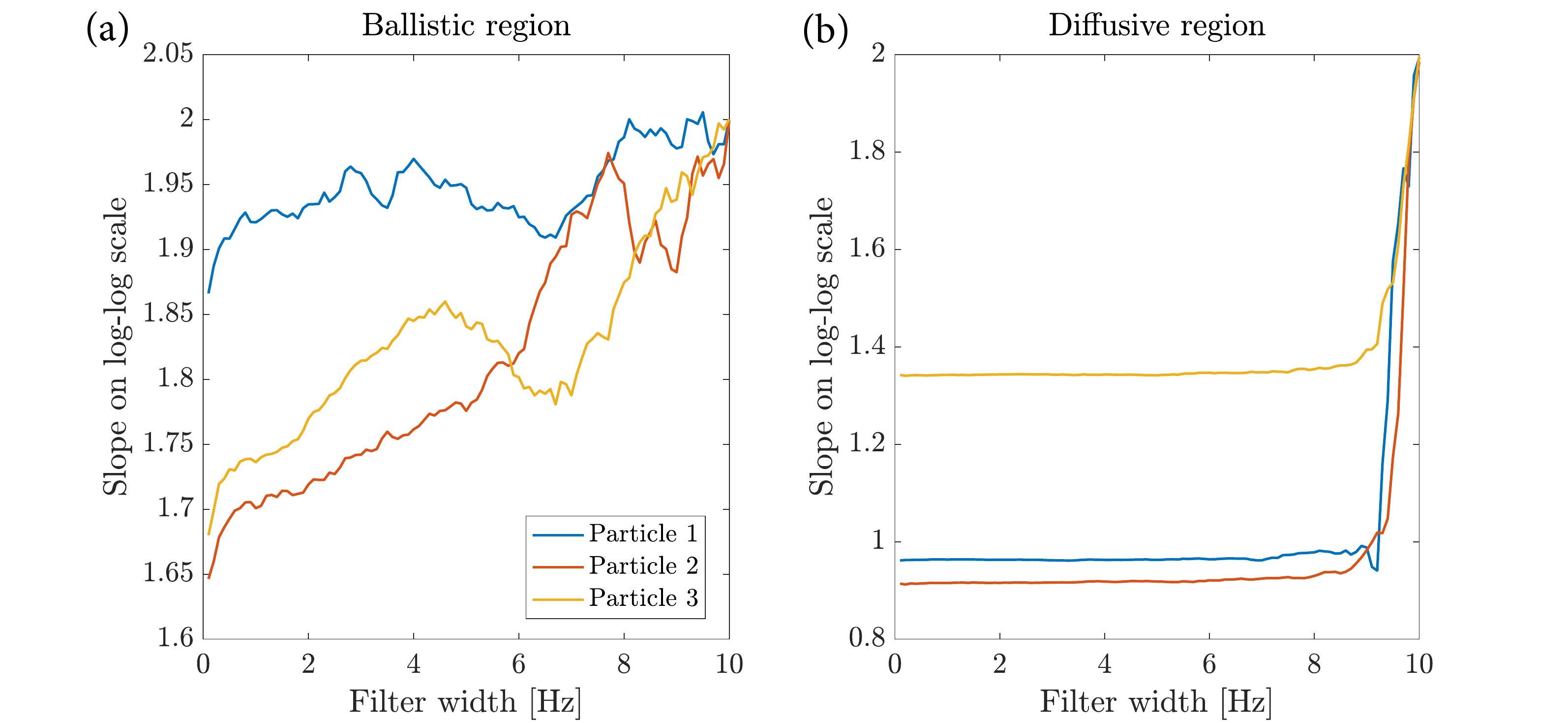}
	\caption{Relationship of fitting parameter
          against filter width for the ballistic (a) and diffusive
          regime (b). The standard error for all plots is $< 0.05$.} 
	\label{fig:params_msd}
\end{figure*}

Fig.~\ref{fig:params_msd} shows how the observed slopes depend on the width of the band-stop filter. 
The effect on the diffusive regime (Fig.~\ref{fig:params_msd}b) is minor. 
The fitted slope is unaffected by filter width below \SI{8}{\Hz}. 
At larger filter widths, the particle trajectory is filtered to a straight line, leading to a slope of two.
The effect of filter width is conversely considerable for the  ballistic regime (Fig.~\ref{fig:params_msd}a). 
The slope is significantly reduced by the presence of the \SI{5}{\Hz} signal for
filter widths $<$ \SI{0.5}{\Hz}. 
With increasing filter width, the slope steadily increases and converges again to a value of two. 
To minimise the effect of the filter, we selected modest filter widths, (\emph{i.e.,} $1 -$\SI{2}{\Hz}).

\begin{table}
\centering
\caption{Fitted slopes of the short and long time scale regimes of Fig.~\ref{fig:X2_loglog}b.}
\label{table:approx_fitting_param_vals}
\begin{tabular}{ccc}
Timescale    & Short & Long \\
\hhline{===}
Particle 1 & $1.91 \pm 0.05$          & $0.96 \pm 0.05$          \\
Particle 2 & $1.70 \pm 0.05$          & $0.91 \pm 0.05$          \\
Particle 3 & $1.69 \pm 0.05$          & $1.34 \pm 0.05$         
\end{tabular}
\end{table} 

\begin{figure}[tbh]
	\centering
	\includegraphics[width=0.45\textwidth]{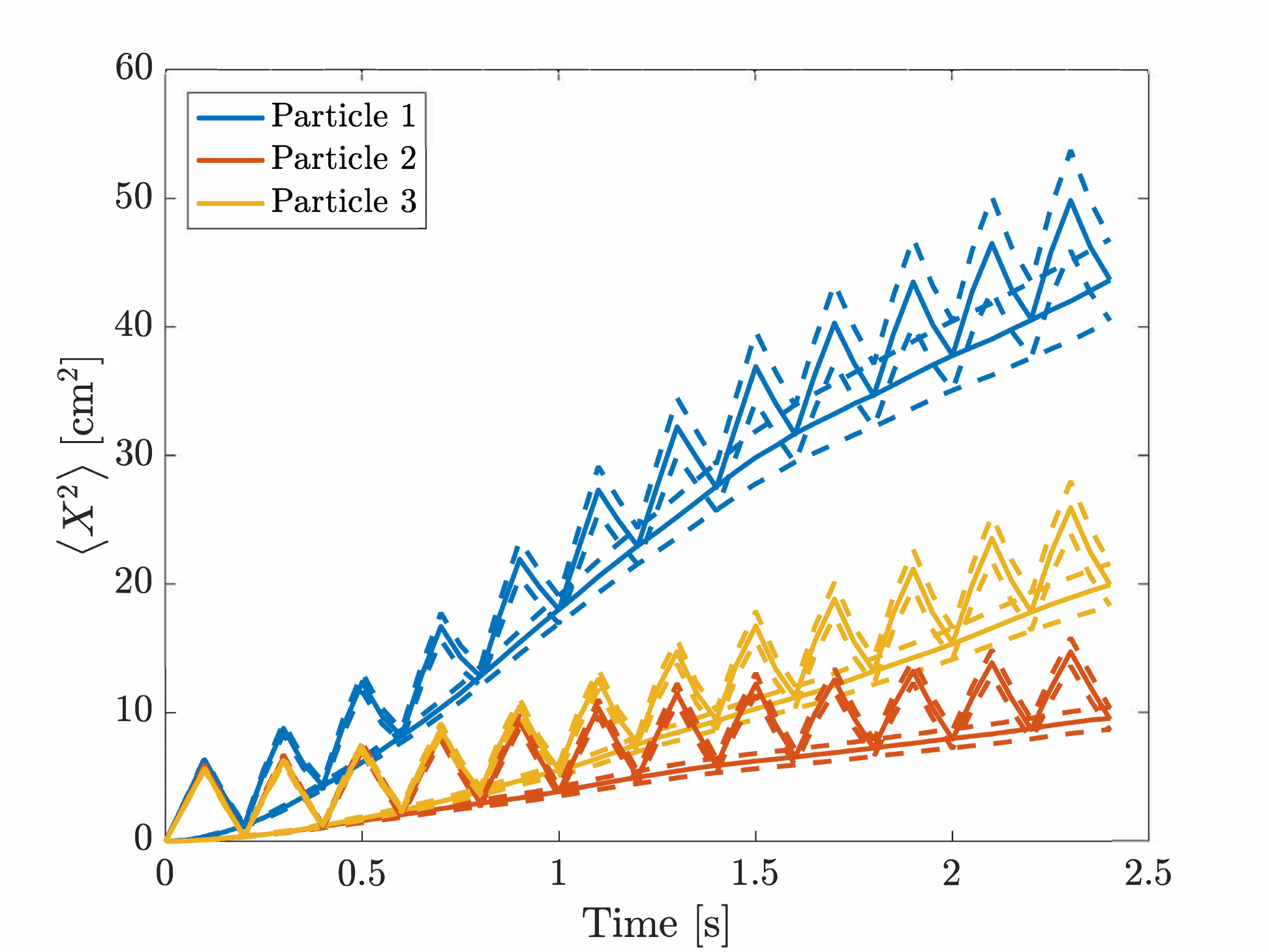}
	\caption{Mean square displacement of filtered and unfiltered
          trajectories for the 3-particle case. Data gathered over 200 trajectories.}
	\label{fig:5Hz_filtering}
\end{figure}

By filtering out the spectral components due to the shaker actuation, we do filter out the \SI{5}{\Hz} components of the natural motion of the particles as well. 
However, the spectrum directly surrounding the \SI{5}{\Hz} peak has an amplitude about $400$ times lower than the low-frequency components of the signal.  
Hence, the disturbing effect on the analysed motion is not significant.  
Additionally, a comparison of the MSD of filtered and unfiltered trajectories (Fig.~\ref{fig:5Hz_filtering}) shows that, in spite of the evident effect of the \SI{5}{\Hz} signal on the particles, 
the baseline signal closely follows the filtered curve. 
This indicates that the filter does not cause the appearance of the ballistic regime in the data. 

The distorted signals and inconsistent behaviour among particles observed in Figs.~6--\ref{fig:params_msd} can be partly explained considering inhomogeneities and selection bias.
Specifically, (1) the particles may not be identical in shape, mass and surface features due to printing tolerances; (2) the reactor surface may be locally uneven, and we observed that different surface properties of the particles or reactor have a marked effect on particle mobility, though they do not affect our general findings; and (3) we manually selected the domains of the ballistic and diffusive regimes.
We would expect the MSD values to converge if the particles were identical.

\subsection{Relative translational diffusion}
Spatial collisions and interactions among particles in self-assembling systems are allowed by the differential motion of the particles. 
To characterise how particles move with respect to one another in the shaken reactor, we define a measure of relative diffusion between two particles $d_\text{rel}$ as the change in distance vector $\mathbf{d}(i)$ with respect to the initial distance $\mathbf{d}(0)$:
\begin{equation}
	d_\text{rel}(i) = \left| \mathbf{d}(i) - \mathbf{d}(0) \right|
\end{equation}
where $\mathbf{d}(i) = [x_1(i)-x_2(i), \ y_1(i)-y_2(i)]$, $x_n$ and $y_n$ being the coordinates of particle $n$ in the shaker frame.
From these values we calculated for the 3-particle case the mean relative square displacements (Fig.~\ref{fig:Diffusion_3spheres_long_relative.pdf}) and relative diffusion coefficients.  
The curves have a similar shape to the absolute diffusion curves. 
The relative diffusion coefficient approximately varies from \SIrange{1}{8}{\square\centi\meter\per\second}. 
Qualitative inspection suggests the relative diffusion equals roughly
the sum of the diffusion of the individual particles, as it might be expected from a Brownian diffusion model~\cite{gillespie2012}, though the relationship was not quantitatively investigated.

\subsection{Rotational diffusion}

We investigated whether the angular orientation of the particles follows a
random walk-like dynamics by observing their angular displacement
over time. 
For this purpose, the values of the angular orientation $\theta$, normally restricted to the
interval $[-\pi,\pi]$, were unwrapped to a continuous, unbounded value.
The result of this conversion is shown in Fig.~D1.   
Similarly to the displacement trajectories, the total angular trajectory was subdivided into shorter trajectories, from which we calculated the mean absolute and the mean squared angular displacement from the initial orientation. 
The latter might be expected to follow \cite{einstein}: 

\begin{equation}\label{angular_diffusion}
	\left< \theta^2 \right> = 2D_\text{r}t
\end{equation}

\noindent where $D_\text{r}$ is the rotational diffusion coefficient, and the scalar factor $2$ follows from the single angular degree of freedom available to the particles. 
The calculated mean absolute angular displacement, shown in Fig.~\ref{fig:Diffusion_3spheres_normal_angle_linear}, appears to grow linearly with time, in contrast to what Eq. \ref{angular_diffusion} predicts. 
Closer observation shows that the particles tend to rotate in a single
direction with approximately fixed rate on a long-term scale; and they do not often undergo random changes of rotation direction typical of random walks.
The different rotational speeds of the particles could possibly be due to the aforementioned physical differences between the individual particles and the unevenness of the reactor surface, so that the momentum is not uniformly transferred from the reactor surface to the particles. 
Moreover, the collisions of the anisotropic particles with the circular border of the reactor are also a source of noise in their angular velocity profiles. 
This confirms that the angular rotation of the particles is heavily biased by the orbital shaking and does not follow a simple diffusive model.

\section{Conclusions}\label{conclusions}

In this paper we studied the kinetics of sliding, centimetre-sized, orbitally-shaken particles by recording and analysing their 2D-constrained motion within a circular bounded space. 
Our results show that the particles possess Rayleigh-like distributed translational velocities in addition to the constant orbital motion components globally forced by the external actuation. 
Orthogonal 1D projections of particle velocity follow a Gaussian-like distribution. 
The mean square displacement of the particles weakly corresponds to a confined random walk model, characterised by the sequence of ballistic, linear and saturating regimes respectively for short, medium and long observation times, which is expected given the presence of hard boundaries to the particle motion.
The relative diffusion coefficient appears approximately equal to the sum of the diffusion coefficient of the particles.
Conversely, the angular particle displacement appears to follow a superdiffusive model. 

Even though, strictly speaking, we did not observe purely ballistic and diffusive motion regimes, we did observe a qualitative trend analogous to that of the molecular realm. 
It is this similarity which we believe may be of interest to the scientific community.  
This similarity moreover arises from a global and periodic actuation of a simple and entirely deterministic system, which makes it rather practical to implement.

The analogy with diffusional kinetics, qualitatively supported by the results of our analysis, is particularly striking when considering that the mechanism underlying the statistics of our particles' motion in the reactor is significantly different from that of, \emph{e.g.,} solutes in a solvent.
In particular, discounting for the impact of air molecules, our particles are not impinged by numerous collisions from other and smaller particles, which conversely defines simple Brownian diffusion~\cite{gillespie2012}.
We hypothesise that the specific motion statistics of our highly diluted particle system may partly arise from properties of the sliding friction between the surfaces of particles and reactor. 
This could be tested by tailoring the surfaces with specific patterns and textures.
Future work will additionally investigate the kinetics of denser 2D orbitally-shaken granular gases of macroscopic sliding particles to develop more effective self-assembly processes.

\begin{acknowledgments}
  The authors would like to thank the good folks at MPI-IS Stuttgart; Thomas Janson, Carsten   Brill and Holger Krause of KIST Europe for experimental support; 
  Per L\"othman, Marc Pichel and Andreas Manz of KIST Europe and Zoey Davidson of MPI-IS Stuttgart for
  fruitful discussions and comments. Marco Dorigo acknowledges the support from the Belgian F.R.S.-FNRS. 
\end{acknowledgments}

\nocite{*} 
\bibliography{aipsamp}

\providecommand{\noopsort}[1]{}\providecommand{\singleletter}[1]{#1}%
\begin{thebibliography}{39}%
\makeatletter
\providecommand \@ifxundefined [1]{%
 \@ifx{#1\undefined}
}%
\providecommand \@ifnum [1]{%
 \ifnum #1\expandafter \@firstoftwo
 \else \expandafter \@secondoftwo
 \fi
}%
\providecommand \@ifx [1]{%
 \ifx #1\expandafter \@firstoftwo
 \else \expandafter \@secondoftwo
 \fi
}%
\providecommand \natexlab [1]{#1}%
\providecommand \enquote  [1]{``#1''}%
\providecommand \bibnamefont  [1]{#1}%
\providecommand \bibfnamefont [1]{#1}%
\providecommand \citenamefont [1]{#1}%
\providecommand \href@noop [0]{\@secondoftwo}%
\providecommand \href [0]{\begingroup \@sanitize@url \@href}%
\providecommand \@href[1]{\@@startlink{#1}\@@href}%
\providecommand \@@href[1]{\endgroup#1\@@endlink}%
\providecommand \@sanitize@url [0]{\catcode `\\12\catcode `\$12\catcode
  `\&12\catcode `\#12\catcode `\^12\catcode `\_12\catcode `\%12\relax}%
\providecommand \@@startlink[1]{}%
\providecommand \@@endlink[0]{}%
\providecommand \url  [0]{\begingroup\@sanitize@url \@url }%
\providecommand \@url [1]{\endgroup\@href {#1}{\urlprefix }}%
\providecommand \urlprefix  [0]{URL }%
\providecommand \Eprint [0]{\href }%
\providecommand \doibase [0]{http://dx.doi.org/}%
\providecommand \selectlanguage [0]{\@gobble}%
\providecommand \bibinfo  [0]{\@secondoftwo}%
\providecommand \bibfield  [0]{\@secondoftwo}%
\providecommand \translation [1]{[#1]}%
\providecommand \BibitemOpen [0]{}%
\providecommand \bibitemStop [0]{}%
\providecommand \bibitemNoStop [0]{.\EOS\space}%
\providecommand \EOS [0]{\spacefactor3000\relax}%
\providecommand \BibitemShut  [1]{\csname bibitem#1\endcsname}%
\let\auto@bib@innerbib\@empty
\bibitem [{\citenamefont {Klug}(1983)}]{Klug}%
  \BibitemOpen
  \bibfield  {author} {\bibinfo {author} {\bibfnamefont {A.}~\bibnamefont
  {Klug}},\ }\href@noop {} {\bibfield  {journal} {\bibinfo  {journal} {Angew.
  Chem. Int. Edit.}\ }\textbf {\bibinfo {volume} {22}},\ \bibinfo {pages} {565}
  (\bibinfo {year} {1983})}\BibitemShut {NoStop}%
\bibitem [{\citenamefont {Whitesides}\ and\ \citenamefont
  {Boncheva}(2002)}]{Whitesides:2002a}%
  \BibitemOpen
  \bibfield  {author} {\bibinfo {author} {\bibfnamefont {G.~M.}\ \bibnamefont
  {Whitesides}}\ and\ \bibinfo {author} {\bibfnamefont {M.}~\bibnamefont
  {Boncheva}},\ }\href@noop {} {\bibfield  {journal} {\bibinfo  {journal} {{P.
  Natl. Acad. Sci. USA}}\ }\textbf {\bibinfo {volume} {99}},\ \bibinfo {pages}
  {4769} (\bibinfo {year} {2002})}\BibitemShut {NoStop}%
\bibitem [{\citenamefont {Whitesides}\ and\ \citenamefont
  {Grzybowski}(2002)}]{Whitesides:2002b}%
  \BibitemOpen
  \bibfield  {author} {\bibinfo {author} {\bibfnamefont {G.~M.}\ \bibnamefont
  {Whitesides}}\ and\ \bibinfo {author} {\bibfnamefont {B.}~\bibnamefont
  {Grzybowski}},\ }\href@noop {} {\bibfield  {journal} {\bibinfo  {journal}
  {Science}\ }\textbf {\bibinfo {volume} {295}},\ \bibinfo {pages} {2418}
  (\bibinfo {year} {2002})}\BibitemShut {NoStop}%
\bibitem [{\citenamefont {Calegari}\ \emph {et~al.}(2014)\citenamefont
  {Calegari}, \citenamefont {Ayuso}, \citenamefont {Trabattoni}, \citenamefont
  {Belshaw}, \citenamefont {De~Camillis}, \citenamefont {Anumula},
  \citenamefont {Frassetto}, \citenamefont {Poletto}, \citenamefont {Palacios},
  \citenamefont {Decleva}, \citenamefont {Greenwood}, \citenamefont
  {Mart{\'\i}n},\ and\ \citenamefont {Nisoli}}]{Calegari2014}%
  \BibitemOpen
  \bibfield  {author} {\bibinfo {author} {\bibfnamefont {F.}~\bibnamefont
  {Calegari}}, \bibinfo {author} {\bibfnamefont {D.}~\bibnamefont {Ayuso}},
  \bibinfo {author} {\bibfnamefont {A.}~\bibnamefont {Trabattoni}}, \bibinfo
  {author} {\bibfnamefont {L.}~\bibnamefont {Belshaw}}, \bibinfo {author}
  {\bibfnamefont {S.}~\bibnamefont {De~Camillis}}, \bibinfo {author}
  {\bibfnamefont {S.}~\bibnamefont {Anumula}}, \bibinfo {author} {\bibfnamefont
  {F.}~\bibnamefont {Frassetto}}, \bibinfo {author} {\bibfnamefont
  {L.}~\bibnamefont {Poletto}}, \bibinfo {author} {\bibfnamefont
  {A.}~\bibnamefont {Palacios}}, \bibinfo {author} {\bibfnamefont
  {P.}~\bibnamefont {Decleva}}, \bibinfo {author} {\bibfnamefont {J.~B.}\
  \bibnamefont {Greenwood}}, \bibinfo {author} {\bibfnamefont {F.}~\bibnamefont
  {Mart{\'\i}n}}, \ and\ \bibinfo {author} {\bibfnamefont {M.}~\bibnamefont
  {Nisoli}},\ }\href {\doibase 10.1126/science.1254061} {\bibfield  {journal}
  {\bibinfo  {journal} {Science}\ }\textbf {\bibinfo {volume} {346}},\ \bibinfo
  {pages} {336} (\bibinfo {year} {2014})}\BibitemShut {NoStop}%
\bibitem [{\citenamefont {Han}\ \emph {et~al.}(2006)\citenamefont {Han},
  \citenamefont {Alsayed}, \citenamefont {Nobili}, \citenamefont {Zhang},
  \citenamefont {Lubensky},\ and\ \citenamefont {Yodh}}]{han2006brownian}%
  \BibitemOpen
  \bibfield  {author} {\bibinfo {author} {\bibfnamefont {Y.}~\bibnamefont
  {Han}}, \bibinfo {author} {\bibfnamefont {A.~M.}\ \bibnamefont {Alsayed}},
  \bibinfo {author} {\bibfnamefont {M.}~\bibnamefont {Nobili}}, \bibinfo
  {author} {\bibfnamefont {J.}~\bibnamefont {Zhang}}, \bibinfo {author}
  {\bibfnamefont {T.~C.}\ \bibnamefont {Lubensky}}, \ and\ \bibinfo {author}
  {\bibfnamefont {A.~G.}\ \bibnamefont {Yodh}},\ }\href@noop {} {\bibfield
  {journal} {\bibinfo  {journal} {Science}\ }\textbf {\bibinfo {volume}
  {314}},\ \bibinfo {pages} {626} (\bibinfo {year} {2006})}\BibitemShut
  {NoStop}%
\bibitem [{\citenamefont {Kheifets}\ \emph {et~al.}(2014)\citenamefont
  {Kheifets}, \citenamefont {Simha}, \citenamefont {Melin}, \citenamefont
  {Li},\ and\ \citenamefont {Raizen}}]{kheifets2014observation}%
  \BibitemOpen
  \bibfield  {author} {\bibinfo {author} {\bibfnamefont {S.}~\bibnamefont
  {Kheifets}}, \bibinfo {author} {\bibfnamefont {A.}~\bibnamefont {Simha}},
  \bibinfo {author} {\bibfnamefont {K.}~\bibnamefont {Melin}}, \bibinfo
  {author} {\bibfnamefont {T.}~\bibnamefont {Li}}, \ and\ \bibinfo {author}
  {\bibfnamefont {M.~G.}\ \bibnamefont {Raizen}},\ }\href@noop {} {\bibfield
  {journal} {\bibinfo  {journal} {Science}\ }\textbf {\bibinfo {volume}
  {343}},\ \bibinfo {pages} {1493} (\bibinfo {year} {2014})}\BibitemShut
  {NoStop}%
\bibitem [{\citenamefont {Huang}\ \emph {et~al.}(2011)\citenamefont {Huang},
  \citenamefont {Chavez}, \citenamefont {Taute}, \citenamefont {Luki{\'c}},
  \citenamefont {Jeney}, \citenamefont {Raizen},\ and\ \citenamefont
  {Florin}}]{huang2011direct}%
  \BibitemOpen
  \bibfield  {author} {\bibinfo {author} {\bibfnamefont {R.}~\bibnamefont
  {Huang}}, \bibinfo {author} {\bibfnamefont {I.}~\bibnamefont {Chavez}},
  \bibinfo {author} {\bibfnamefont {K.~M.}\ \bibnamefont {Taute}}, \bibinfo
  {author} {\bibfnamefont {B.}~\bibnamefont {Luki{\'c}}}, \bibinfo {author}
  {\bibfnamefont {S.}~\bibnamefont {Jeney}}, \bibinfo {author} {\bibfnamefont
  {M.~G.}\ \bibnamefont {Raizen}}, \ and\ \bibinfo {author} {\bibfnamefont
  {E.-L.}\ \bibnamefont {Florin}},\ }\href@noop {} {\bibfield  {journal}
  {\bibinfo  {journal} {Nat. Phys.}\ }\textbf {\bibinfo {volume} {7}},\
  \bibinfo {pages} {576} (\bibinfo {year} {2011})}\BibitemShut {NoStop}%
\bibitem [{\citenamefont {Loudet}\ \emph {et~al.}(2004)\citenamefont {Loudet},
  \citenamefont {Hanusse},\ and\ \citenamefont {Poulin}}]{loudet2004stokes}%
  \BibitemOpen
  \bibfield  {author} {\bibinfo {author} {\bibfnamefont {J.~C.}\ \bibnamefont
  {Loudet}}, \bibinfo {author} {\bibfnamefont {P.}~\bibnamefont {Hanusse}}, \
  and\ \bibinfo {author} {\bibfnamefont {P.}~\bibnamefont {Poulin}},\
  }\href@noop {} {\bibfield  {journal} {\bibinfo  {journal} {Science}\ }\textbf
  {\bibinfo {volume} {306}},\ \bibinfo {pages} {1525} (\bibinfo {year}
  {2004})}\BibitemShut {NoStop}%
\bibitem [{\citenamefont {Liu}\ and\ \citenamefont
  {B{\"o}ker}(2016)}]{liu2016measuring}%
  \BibitemOpen
  \bibfield  {author} {\bibinfo {author} {\bibfnamefont {B.}~\bibnamefont
  {Liu}}\ and\ \bibinfo {author} {\bibfnamefont {A.}~\bibnamefont
  {B{\"o}ker}},\ }\href@noop {} {\bibfield  {journal} {\bibinfo  {journal}
  {Soft matter}\ }\textbf {\bibinfo {volume} {12}},\ \bibinfo {pages} {6033}
  (\bibinfo {year} {2016})}\BibitemShut {NoStop}%
\bibitem [{\citenamefont {Chakrabarty}\ \emph {et~al.}(2014)\citenamefont
  {Chakrabarty}, \citenamefont {Konya}, \citenamefont {Wang}, \citenamefont
  {Selinger}, \citenamefont {Sun},\ and\ \citenamefont
  {Wei}}]{chakrabarty2014brownian}%
  \BibitemOpen
  \bibfield  {author} {\bibinfo {author} {\bibfnamefont {A.}~\bibnamefont
  {Chakrabarty}}, \bibinfo {author} {\bibfnamefont {A.}~\bibnamefont {Konya}},
  \bibinfo {author} {\bibfnamefont {F.}~\bibnamefont {Wang}}, \bibinfo {author}
  {\bibfnamefont {J.~V.}\ \bibnamefont {Selinger}}, \bibinfo {author}
  {\bibfnamefont {K.}~\bibnamefont {Sun}}, \ and\ \bibinfo {author}
  {\bibfnamefont {Q.-H.}\ \bibnamefont {Wei}},\ }\href@noop {} {\bibfield
  {journal} {\bibinfo  {journal} {Langmuir}\ }\textbf {\bibinfo {volume}
  {30}},\ \bibinfo {pages} {13844} (\bibinfo {year} {2014})}\BibitemShut
  {NoStop}%
\bibitem [{\citenamefont {Roy}\ \emph {et~al.}(2016)\citenamefont {Roy},
  \citenamefont {Mondal}, \citenamefont {Bera},\ and\ \citenamefont
  {Banerjee}}]{roy2016using}%
  \BibitemOpen
  \bibfield  {author} {\bibinfo {author} {\bibfnamefont {B.}~\bibnamefont
  {Roy}}, \bibinfo {author} {\bibfnamefont {A.}~\bibnamefont {Mondal}},
  \bibinfo {author} {\bibfnamefont {S.~K.}\ \bibnamefont {Bera}}, \ and\
  \bibinfo {author} {\bibfnamefont {A.}~\bibnamefont {Banerjee}},\ }\href@noop
  {} {\bibfield  {journal} {\bibinfo  {journal} {Soft Matter}\ }\textbf
  {\bibinfo {volume} {12}},\ \bibinfo {pages} {5077} (\bibinfo {year}
  {2016})}\BibitemShut {NoStop}%
\bibitem [{\citenamefont {Ilievski}\ \emph {et~al.}(2011)\citenamefont
  {Ilievski}, \citenamefont {Mani}, \citenamefont {Whitesides},\ and\
  \citenamefont {Brenner}}]{Ilievski}%
  \BibitemOpen
  \bibfield  {author} {\bibinfo {author} {\bibfnamefont {F.}~\bibnamefont
  {Ilievski}}, \bibinfo {author} {\bibfnamefont {M.}~\bibnamefont {Mani}},
  \bibinfo {author} {\bibfnamefont {G.~M.}\ \bibnamefont {Whitesides}}, \ and\
  \bibinfo {author} {\bibfnamefont {M.~P.}\ \bibnamefont {Brenner}},\ }\href
  {\doibase 10.1103/PhysRevE.83.017301} {\bibfield  {journal} {\bibinfo
  {journal} {Phys. Rev. E}\ }\textbf {\bibinfo {volume} {83}},\ \bibinfo
  {pages} {017301} (\bibinfo {year} {2011})}\BibitemShut {NoStop}%
\bibitem [{\citenamefont {Hageman}\ \emph {et~al.}(2018)\citenamefont
  {Hageman}, \citenamefont {L\"{o}thman}, \citenamefont {Dirnberger},
  \citenamefont {Elwenspoek}, \citenamefont {Manz},\ and\ \citenamefont
  {Abelmann}}]{hageman2018MacroEquivalence}%
  \BibitemOpen
  \bibfield  {author} {\bibinfo {author} {\bibfnamefont {T.~A.~G.}\
  \bibnamefont {Hageman}}, \bibinfo {author} {\bibfnamefont {P.~A.}\
  \bibnamefont {L\"{o}thman}}, \bibinfo {author} {\bibfnamefont
  {M.}~\bibnamefont {Dirnberger}}, \bibinfo {author} {\bibfnamefont {M.~C.}\
  \bibnamefont {Elwenspoek}}, \bibinfo {author} {\bibfnamefont
  {A.}~\bibnamefont {Manz}}, \ and\ \bibinfo {author} {\bibfnamefont
  {L.}~\bibnamefont {Abelmann}},\ }\href@noop {} {\bibfield  {journal}
  {\bibinfo  {journal} {J. Appl. Phys.}\ }\textbf {\bibinfo {volume} {123}},\
  \bibinfo {pages} {024901} (\bibinfo {year} {2018})}\BibitemShut {NoStop}%
\bibitem [{\citenamefont {Bian}\ \emph {et~al.}(2016)\citenamefont {Bian},
  \citenamefont {Kim},\ and\ \citenamefont {Karniadakis}}]{bian2016111}%
  \BibitemOpen
  \bibfield  {author} {\bibinfo {author} {\bibfnamefont {X.}~\bibnamefont
  {Bian}}, \bibinfo {author} {\bibfnamefont {C.}~\bibnamefont {Kim}}, \ and\
  \bibinfo {author} {\bibfnamefont {G.~E.}\ \bibnamefont {Karniadakis}},\
  }\href@noop {} {\bibfield  {journal} {\bibinfo  {journal} {Soft Matter}\
  }\textbf {\bibinfo {volume} {12}},\ \bibinfo {pages} {6331} (\bibinfo {year}
  {2016})}\BibitemShut {NoStop}%
\bibitem [{\citenamefont {Grzybowski}\ \emph {et~al.}(2002)\citenamefont
  {Grzybowski}, \citenamefont {Stone},\ and\ \citenamefont
  {Whitesides}}]{Grzybowski}%
  \BibitemOpen
  \bibfield  {author} {\bibinfo {author} {\bibfnamefont {B.~A.}\ \bibnamefont
  {Grzybowski}}, \bibinfo {author} {\bibfnamefont {H.~A.}\ \bibnamefont
  {Stone}}, \ and\ \bibinfo {author} {\bibfnamefont {G.~M.}\ \bibnamefont
  {Whitesides}},\ }\href@noop {} {\bibfield  {journal} {\bibinfo  {journal}
  {{P. Natl. Acad. Sci. USA}}\ }\textbf {\bibinfo {volume} {99}},\ \bibinfo
  {pages} {4147} (\bibinfo {year} {2002})}\BibitemShut {NoStop}%
\bibitem [{\citenamefont {Hosokawa}\ \emph {et~al.}(1994)\citenamefont
  {Hosokawa}, \citenamefont {Shimoyama},\ and\ \citenamefont
  {Miura}}]{Hosokawa1994}%
  \BibitemOpen
  \bibfield  {author} {\bibinfo {author} {\bibfnamefont {K.}~\bibnamefont
  {Hosokawa}}, \bibinfo {author} {\bibfnamefont {I.}~\bibnamefont {Shimoyama}},
  \ and\ \bibinfo {author} {\bibfnamefont {H.}~\bibnamefont {Miura}},\
  }\href@noop {} {\bibfield  {journal} {\bibinfo  {journal} {Artif. Life}\
  }\textbf {\bibinfo {volume} {1}},\ \bibinfo {pages} {413} (\bibinfo {year}
  {1994})}\BibitemShut {NoStop}%
\bibitem [{\citenamefont {Miyashita}\ \emph {et~al.}(2011)\citenamefont
  {Miyashita}, \citenamefont {G\"{o}ldi},\ and\ \citenamefont
  {Pfeifer}}]{Miyashita01042011}%
  \BibitemOpen
  \bibfield  {author} {\bibinfo {author} {\bibfnamefont {S.}~\bibnamefont
  {Miyashita}}, \bibinfo {author} {\bibfnamefont {M.}~\bibnamefont
  {G\"{o}ldi}}, \ and\ \bibinfo {author} {\bibfnamefont {R.}~\bibnamefont
  {Pfeifer}},\ }\href@noop {} {\bibfield  {journal} {\bibinfo  {journal} {Int.
  J. Robot. Res.}\ }\textbf {\bibinfo {volume} {30}},\ \bibinfo {pages} {627}
  (\bibinfo {year} {2011})}\BibitemShut {NoStop}%
\bibitem [{\citenamefont {Ipparthi}\ \emph {et~al.}(2017)\citenamefont
  {Ipparthi}, \citenamefont {Winslow}, \citenamefont {Sitti}, \citenamefont
  {Dorigo},\ and\ \citenamefont {Mastrangeli}}]{ipparthi2017yield}%
  \BibitemOpen
  \bibfield  {author} {\bibinfo {author} {\bibfnamefont {D.}~\bibnamefont
  {Ipparthi}}, \bibinfo {author} {\bibfnamefont {A.}~\bibnamefont {Winslow}},
  \bibinfo {author} {\bibfnamefont {M.}~\bibnamefont {Sitti}}, \bibinfo
  {author} {\bibfnamefont {M.}~\bibnamefont {Dorigo}}, \ and\ \bibinfo {author}
  {\bibfnamefont {M.}~\bibnamefont {Mastrangeli}},\ }\href@noop {} {\bibfield
  {journal} {\bibinfo  {journal} {Soft Matter}\ }\textbf {\bibinfo {volume}
  {13}},\ \bibinfo {pages} {7595} (\bibinfo {year} {2017})}\BibitemShut
  {NoStop}%
\bibitem [{\citenamefont {Gillespie}(2007)}]{Gillespie_2007}%
  \BibitemOpen
  \bibfield  {author} {\bibinfo {author} {\bibfnamefont {D.~T.}\ \bibnamefont
  {Gillespie}},\ }\href@noop {} {\bibfield  {journal} {\bibinfo  {journal}
  {Annu. Rev. Phys. Chem.}\ }\textbf {\bibinfo {volume} {58}},\ \bibinfo
  {pages} {35} (\bibinfo {year} {2007})}\BibitemShut {NoStop}%
\bibitem [{\citenamefont {Ojha}\ \emph {et~al.}(2004)\citenamefont {Ojha},
  \citenamefont {Lemieux}, \citenamefont {Dixon}, \citenamefont {Liu},\ and\
  \citenamefont {Durian}}]{ojha2004}%
  \BibitemOpen
  \bibfield  {author} {\bibinfo {author} {\bibfnamefont {R.}~\bibnamefont
  {Ojha}}, \bibinfo {author} {\bibfnamefont {P.}~\bibnamefont {Lemieux}},
  \bibinfo {author} {\bibfnamefont {P.}~\bibnamefont {Dixon}}, \bibinfo
  {author} {\bibfnamefont {A.}~\bibnamefont {Liu}}, \ and\ \bibinfo {author}
  {\bibfnamefont {D.}~\bibnamefont {Durian}},\ }\href@noop {} {\bibfield
  {journal} {\bibinfo  {journal} {Nature}\ }\textbf {\bibinfo {volume} {427}},\
  \bibinfo {pages} {521} (\bibinfo {year} {2004})}\BibitemShut {NoStop}%
\bibitem [{\citenamefont {Kudrolli}(2004)}]{kudrolli2004}%
  \BibitemOpen
  \bibfield  {author} {\bibinfo {author} {\bibfnamefont {A.}~\bibnamefont
  {Kudrolli}},\ }\href@noop {} {\bibfield  {journal} {\bibinfo  {journal} {Rep.
  Prog. Phys.}\ }\textbf {\bibinfo {volume} {67}},\ \bibinfo {pages} {209}
  (\bibinfo {year} {2004})}\BibitemShut {NoStop}%
\bibitem [{\citenamefont {Kumar}\ \emph {et~al.}(2015)\citenamefont {Kumar},
  \citenamefont {Nitsure}, \citenamefont {Bhattacharya},\ and\ \citenamefont
  {Ghosh}}]{Kumar2015}%
  \BibitemOpen
  \bibfield  {author} {\bibinfo {author} {\bibfnamefont {D.}~\bibnamefont
  {Kumar}}, \bibinfo {author} {\bibfnamefont {N.}~\bibnamefont {Nitsure}},
  \bibinfo {author} {\bibfnamefont {S.}~\bibnamefont {Bhattacharya}}, \ and\
  \bibinfo {author} {\bibfnamefont {S.}~\bibnamefont {Ghosh}},\ }\href@noop {}
  {\bibfield  {journal} {\bibinfo  {journal} {Proc. Natl. Acad. Sci. USA}\
  }\textbf {\bibinfo {volume} {112}},\ \bibinfo {pages} {11443} (\bibinfo
  {year} {2015})}\BibitemShut {NoStop}%
\bibitem [{\citenamefont {Cademartiri}\ \emph {et~al.}(2012)\citenamefont
  {Cademartiri}, \citenamefont {Stan}, \citenamefont {Tran}, \citenamefont
  {Wu}, \citenamefont {Friar}, \citenamefont {Vulis}, \citenamefont {Clark},
  \citenamefont {Tricard},\ and\ \citenamefont {Whitesides}}]{Cademartiri2012}%
  \BibitemOpen
  \bibfield  {author} {\bibinfo {author} {\bibfnamefont {R.}~\bibnamefont
  {Cademartiri}}, \bibinfo {author} {\bibfnamefont {C.~A.}\ \bibnamefont
  {Stan}}, \bibinfo {author} {\bibfnamefont {V.~M.}\ \bibnamefont {Tran}},
  \bibinfo {author} {\bibfnamefont {E.}~\bibnamefont {Wu}}, \bibinfo {author}
  {\bibfnamefont {L.}~\bibnamefont {Friar}}, \bibinfo {author} {\bibfnamefont
  {D.}~\bibnamefont {Vulis}}, \bibinfo {author} {\bibfnamefont {L.~W.}\
  \bibnamefont {Clark}}, \bibinfo {author} {\bibfnamefont {S.}~\bibnamefont
  {Tricard}}, \ and\ \bibinfo {author} {\bibfnamefont {G.~M.}\ \bibnamefont
  {Whitesides}},\ }\href@noop {} {\bibfield  {journal} {\bibinfo  {journal}
  {Soft Matter}\ }\textbf {\bibinfo {volume} {8}},\ \bibinfo {pages} {9771}
  (\bibinfo {year} {2012})}\BibitemShut {NoStop}%
\bibitem [{\citenamefont {Tricard}\ \emph {et~al.}(2013)\citenamefont
  {Tricard}, \citenamefont {Stan}, \citenamefont {Shakhnovich},\ and\
  \citenamefont {Whitesides}}]{Tricard2013}%
  \BibitemOpen
  \bibfield  {author} {\bibinfo {author} {\bibfnamefont {S.}~\bibnamefont
  {Tricard}}, \bibinfo {author} {\bibfnamefont {C.~A.}\ \bibnamefont {Stan}},
  \bibinfo {author} {\bibfnamefont {E.~I.}\ \bibnamefont {Shakhnovich}}, \ and\
  \bibinfo {author} {\bibfnamefont {G.~M.}\ \bibnamefont {Whitesides}},\
  }\href@noop {} {\bibfield  {journal} {\bibinfo  {journal} {Soft Matter}\
  }\textbf {\bibinfo {volume} {9}},\ \bibinfo {pages} {4480} (\bibinfo {year}
  {2013})}\BibitemShut {NoStop}%
\bibitem [{\citenamefont {Tricard}\ \emph {et~al.}(2015)\citenamefont
  {Tricard}, \citenamefont {Shepherd}, \citenamefont {Stan}, \citenamefont
  {Snyder}, \citenamefont {Cademartiri}, \citenamefont {Zhu}, \citenamefont
  {Aranson}, \citenamefont {Shakhnovich},\ and\ \citenamefont
  {Whitesides}}]{Tricard2015}%
  \BibitemOpen
  \bibfield  {author} {\bibinfo {author} {\bibfnamefont {S.}~\bibnamefont
  {Tricard}}, \bibinfo {author} {\bibfnamefont {R.~F.}\ \bibnamefont
  {Shepherd}}, \bibinfo {author} {\bibfnamefont {C.~A.}\ \bibnamefont {Stan}},
  \bibinfo {author} {\bibfnamefont {P.~W.}\ \bibnamefont {Snyder}}, \bibinfo
  {author} {\bibfnamefont {R.}~\bibnamefont {Cademartiri}}, \bibinfo {author}
  {\bibfnamefont {D.}~\bibnamefont {Zhu}}, \bibinfo {author} {\bibfnamefont
  {I.~S.}\ \bibnamefont {Aranson}}, \bibinfo {author} {\bibfnamefont {E.~I.}\
  \bibnamefont {Shakhnovich}}, \ and\ \bibinfo {author} {\bibfnamefont {G.~M.}\
  \bibnamefont {Whitesides}},\ }\href@noop {} {\bibfield  {journal} {\bibinfo
  {journal} {ChemPlusChem}\ }\textbf {\bibinfo {volume} {80}},\ \bibinfo
  {pages} {37} (\bibinfo {year} {2015})}\BibitemShut {NoStop}%
\bibitem [{\citenamefont {Hacohen}\ \emph {et~al.}(2015)\citenamefont
  {Hacohen}, \citenamefont {Hanniel}, \citenamefont {Nikulshin}, \citenamefont
  {Wolfus}, \citenamefont {Abu-Horowitz},\ and\ \citenamefont
  {Bachelet}}]{Hacohen2015}%
  \BibitemOpen
  \bibfield  {author} {\bibinfo {author} {\bibfnamefont {A.}~\bibnamefont
  {Hacohen}}, \bibinfo {author} {\bibfnamefont {I.}~\bibnamefont {Hanniel}},
  \bibinfo {author} {\bibfnamefont {Y.}~\bibnamefont {Nikulshin}}, \bibinfo
  {author} {\bibfnamefont {S.}~\bibnamefont {Wolfus}}, \bibinfo {author}
  {\bibfnamefont {A.}~\bibnamefont {Abu-Horowitz}}, \ and\ \bibinfo {author}
  {\bibfnamefont {I.}~\bibnamefont {Bachelet}},\ }\href@noop {} {\bibfield
  {journal} {\bibinfo  {journal} {Sci. Rep.}\ }\textbf {\bibinfo {volume} {5}}
  (\bibinfo {year} {2015})}\BibitemShut {NoStop}%
\bibitem [{\citenamefont {Bhalla}\ \emph {et~al.}(2014)\citenamefont {Bhalla},
  \citenamefont {Ipparthi}, \citenamefont {Klemp},\ and\ \citenamefont
  {Dorigo}}]{Bhalla2014}%
  \BibitemOpen
  \bibfield  {author} {\bibinfo {author} {\bibfnamefont {N.}~\bibnamefont
  {Bhalla}}, \bibinfo {author} {\bibfnamefont {D.}~\bibnamefont {Ipparthi}},
  \bibinfo {author} {\bibfnamefont {E.}~\bibnamefont {Klemp}}, \ and\ \bibinfo
  {author} {\bibfnamefont {M.}~\bibnamefont {Dorigo}},\ }in\ \href@noop {}
  {\emph {\bibinfo {booktitle} {{13th Int. Conf. on Parallel Problem Solving
  from Nature (PPSN 2014)}}}}\ (\bibinfo {organization} {Springer},\ \bibinfo
  {year} {2014})\ pp.\ \bibinfo {pages} {751--760}\BibitemShut {NoStop}%
\bibitem [{\citenamefont {Gr\"unwald}\ \emph {et~al.}(2016)\citenamefont
  {Gr\"unwald}, \citenamefont {Tricard}, \citenamefont {Whitesides},\ and\
  \citenamefont {Geissler}}]{Grunwald2016}%
  \BibitemOpen
  \bibfield  {author} {\bibinfo {author} {\bibfnamefont {M.}~\bibnamefont
  {Gr\"unwald}}, \bibinfo {author} {\bibfnamefont {S.}~\bibnamefont {Tricard}},
  \bibinfo {author} {\bibfnamefont {G.~M.}\ \bibnamefont {Whitesides}}, \ and\
  \bibinfo {author} {\bibfnamefont {P.~L.}\ \bibnamefont {Geissler}},\
  }\href@noop {} {\bibfield  {journal} {\bibinfo  {journal} {Soft Matter}\
  }\textbf {\bibinfo {volume} {12}},\ \bibinfo {pages} {1517} (\bibinfo {year}
  {2016})}\BibitemShut {NoStop}%
\bibitem [{\citenamefont {Scherer}\ \emph {et~al.}(2000)\citenamefont
  {Scherer}, \citenamefont {K\"otter}, \citenamefont {Markus}, \citenamefont
  {Goles},\ and\ \citenamefont {Rehberg}}]{Scherer2000}%
  \BibitemOpen
  \bibfield  {author} {\bibinfo {author} {\bibfnamefont {M.~A.}\ \bibnamefont
  {Scherer}}, \bibinfo {author} {\bibfnamefont {K.}~\bibnamefont {K\"otter}},
  \bibinfo {author} {\bibfnamefont {M.}~\bibnamefont {Markus}}, \bibinfo
  {author} {\bibfnamefont {E.}~\bibnamefont {Goles}}, \ and\ \bibinfo {author}
  {\bibfnamefont {I.}~\bibnamefont {Rehberg}},\ }\href {\doibase
  10.1103/PhysRevE.61.4069} {\bibfield  {journal} {\bibinfo  {journal} {Phys.
  Rev. E}\ }\textbf {\bibinfo {volume} {61}},\ \bibinfo {pages} {4069}
  (\bibinfo {year} {2000})}\BibitemShut {NoStop}%
\bibitem [{\citenamefont {Gillespie}\ and\ \citenamefont
  {Seitaridou}(2012)}]{gillespie2012}%
  \BibitemOpen
  \bibfield  {author} {\bibinfo {author} {\bibfnamefont {D.~T.}\ \bibnamefont
  {Gillespie}}\ and\ \bibinfo {author} {\bibfnamefont {E.}~\bibnamefont
  {Seitaridou}},\ }\href@noop {} {\emph {\bibinfo {title} {Simple Brownian
  diffusion: an introduction to the standard theoretical models}}}\ (\bibinfo
  {publisher} {Oxford University Press},\ \bibinfo {year} {2012})\BibitemShut
  {NoStop}%
\bibitem [{\citenamefont {Einstein}(1905)}]{einstein}%
  \BibitemOpen
  \bibfield  {author} {\bibinfo {author} {\bibfnamefont {A.}~\bibnamefont
  {Einstein}},\ }\href@noop {} {\bibfield  {journal} {\bibinfo  {journal} {Ann.
  Phys.}\ }\textbf {\bibinfo {volume} {17}} (\bibinfo {year}
  {1905})}\BibitemShut {NoStop}%
\bibitem [{\citenamefont {{Smoluchowski}}(1916)}]{smoluchowski}%
  \BibitemOpen
  \bibfield  {author} {\bibinfo {author} {\bibfnamefont {M.~V.}\ \bibnamefont
  {{Smoluchowski}}},\ }\href@noop {} {\bibfield  {journal} {\bibinfo  {journal}
  {Physik. Zeit.}\ }\textbf {\bibinfo {volume} {17}},\ \bibinfo {pages} {557}
  (\bibinfo {year} {1916})}\BibitemShut {NoStop}%
\bibitem [{\citenamefont {Langevin}(1908)}]{Langevin1908}%
  \BibitemOpen
  \bibfield  {author} {\bibinfo {author} {\bibfnamefont {P.}~\bibnamefont
  {Langevin}},\ }\href@noop {} {\bibfield  {journal} {\bibinfo  {journal} {C.
  R. Acad. Sci.}\ }\textbf {\bibinfo {volume} {146}},\ \bibinfo {pages} {530}
  (\bibinfo {year} {1908})}\BibitemShut {NoStop}%
\bibitem [{\citenamefont {Cooksy}(2014)}]{cooksy2014}%
  \BibitemOpen
  \bibfield  {author} {\bibinfo {author} {\bibfnamefont {A.}~\bibnamefont
  {Cooksy}},\ }\href@noop {} {\emph {\bibinfo {title} {Physical Chemistry:
  Thermodynamics, Statistical Mechanics \& Kinetics}}}\ (\bibinfo  {publisher}
  {Pearson},\ \bibinfo {year} {2014})\BibitemShut {NoStop}%
\bibitem [{\citenamefont {Berg}(1993)}]{berg1993}%
  \BibitemOpen
  \bibfield  {author} {\bibinfo {author} {\bibfnamefont {H.~C.}\ \bibnamefont
  {Berg}},\ }\href@noop {} {\emph {\bibinfo {title} {Random walks in
  biology}}}\ (\bibinfo  {publisher} {Princeton University Press},\ \bibinfo
  {year} {1993})\BibitemShut {NoStop}%
\bibitem [{Note2()}]{Note2}%
  \BibitemOpen
  \bibinfo {note} {Figs.~A1, B1, C1 and D1 are available in the
  Appendix.}\BibitemShut {Stop}%
\bibitem [{\citenamefont {Press}\ \emph {et~al.}(1992)\citenamefont {Press},
  \citenamefont {Teukolsky}, \citenamefont {Vetterling},\ and\ \citenamefont
  {Flannery}}]{Press1992}%
  \BibitemOpen
  \bibfield  {author} {\bibinfo {author} {\bibfnamefont {W.~H.}\ \bibnamefont
  {Press}}, \bibinfo {author} {\bibfnamefont {S.~A.}\ \bibnamefont
  {Teukolsky}}, \bibinfo {author} {\bibfnamefont {W.~T.}\ \bibnamefont
  {Vetterling}}, \ and\ \bibinfo {author} {\bibfnamefont {B.~P.}\ \bibnamefont
  {Flannery}},\ }\href@noop {} {\emph {\bibinfo {title} {Numerical Recipes in C
  (2{nd} Ed.): The Art of Scientific Computing}}}\ (\bibinfo  {publisher}
  {Cambridge University Press},\ \bibinfo {address} {New York, NY, USA},\
  \bibinfo {year} {1992})\BibitemShut {NoStop}%
\bibitem [{\citenamefont {Perrin}(1909)}]{perrin}%
  \BibitemOpen
  \bibfield  {author} {\bibinfo {author} {\bibfnamefont {J.}~\bibnamefont
  {Perrin}},\ }\href@noop {} {\bibfield  {journal} {\bibinfo  {journal} {Ann.
  Chim. Phys.}\ }\textbf {\bibinfo {volume} {18}},\ \bibinfo {pages} {5}
  (\bibinfo {year} {1909})}\BibitemShut {NoStop}%
\bibitem [{Note1()}]{Note1}%
  \BibitemOpen
  \bibinfo {note} {\protect \S ~Current address: Electronic Components,
  Technology and Materials, Department of Microelectronics, Delft University of
  Technology, Delft, The Netherlands}\BibitemShut {NoStop}%
\end{thebibliography}%

\appendix
\onecolumngrid

\clearpage
\section{Cumulative 2D velocity distribution}\label{app:cumulative_distribution}

\renewcommand\thefigure{A\arabic{figure}}    
\setcounter{figure}{0}  
\begin{figure}[h]
	\centering	\includegraphics[width=0.5\textwidth]{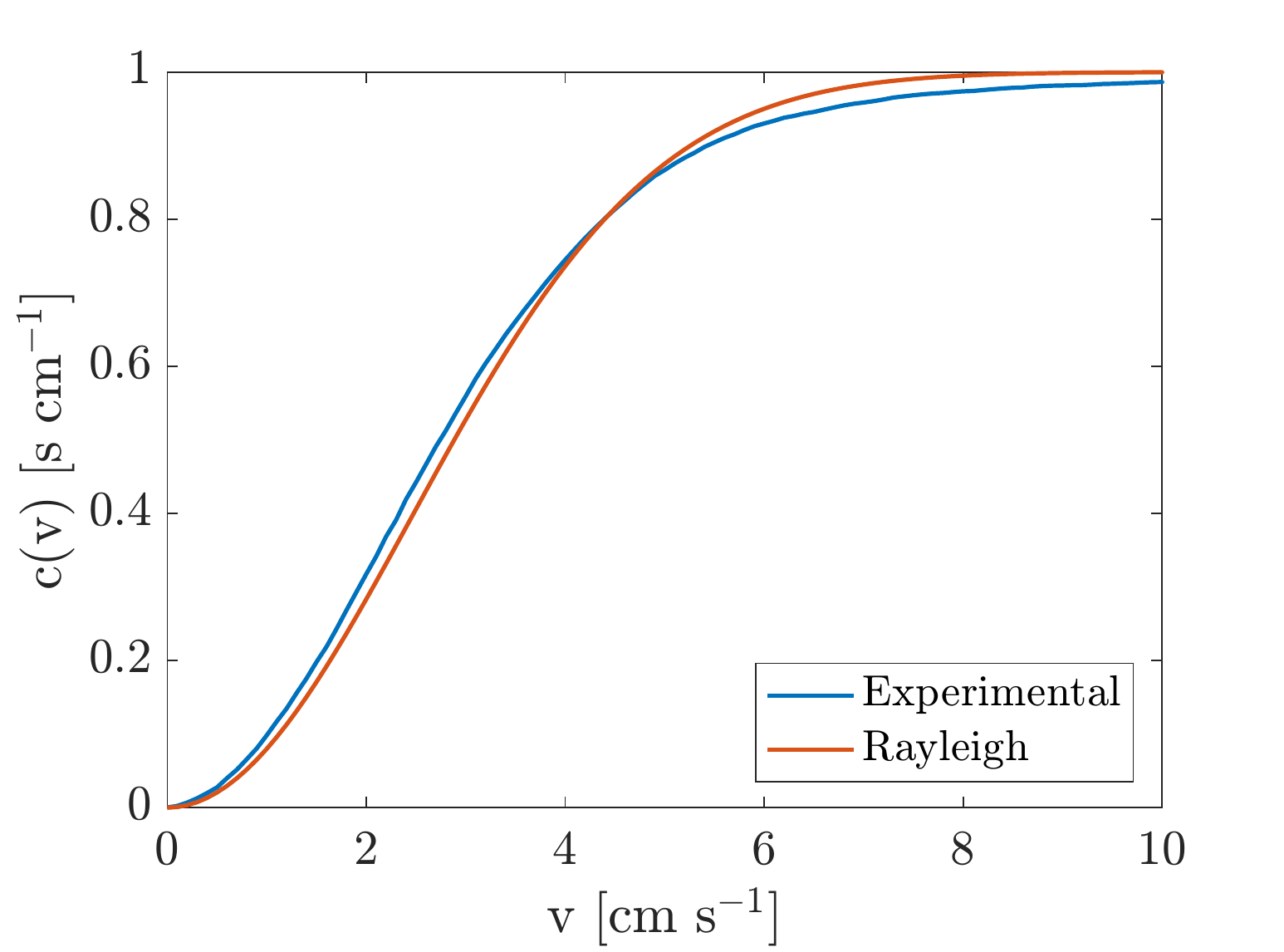}
	\caption{Normalised cumulative distribution of the measured and filtered 2D velocity of a single particle (see Fig.~3) and the fitted cumulative Rayleigh distribution.}
	\label{fig:cumulative_distribution}
\end{figure}

\section{Spatial particle velocity distribution}
\renewcommand\thefigure{B\arabic{figure}}    
\setcounter{figure}{0}  

\begin{figure*}[h]
  \centering
  \begin{subfigure}[b]{0.48\textwidth}
    \includegraphics[width=\textwidth]{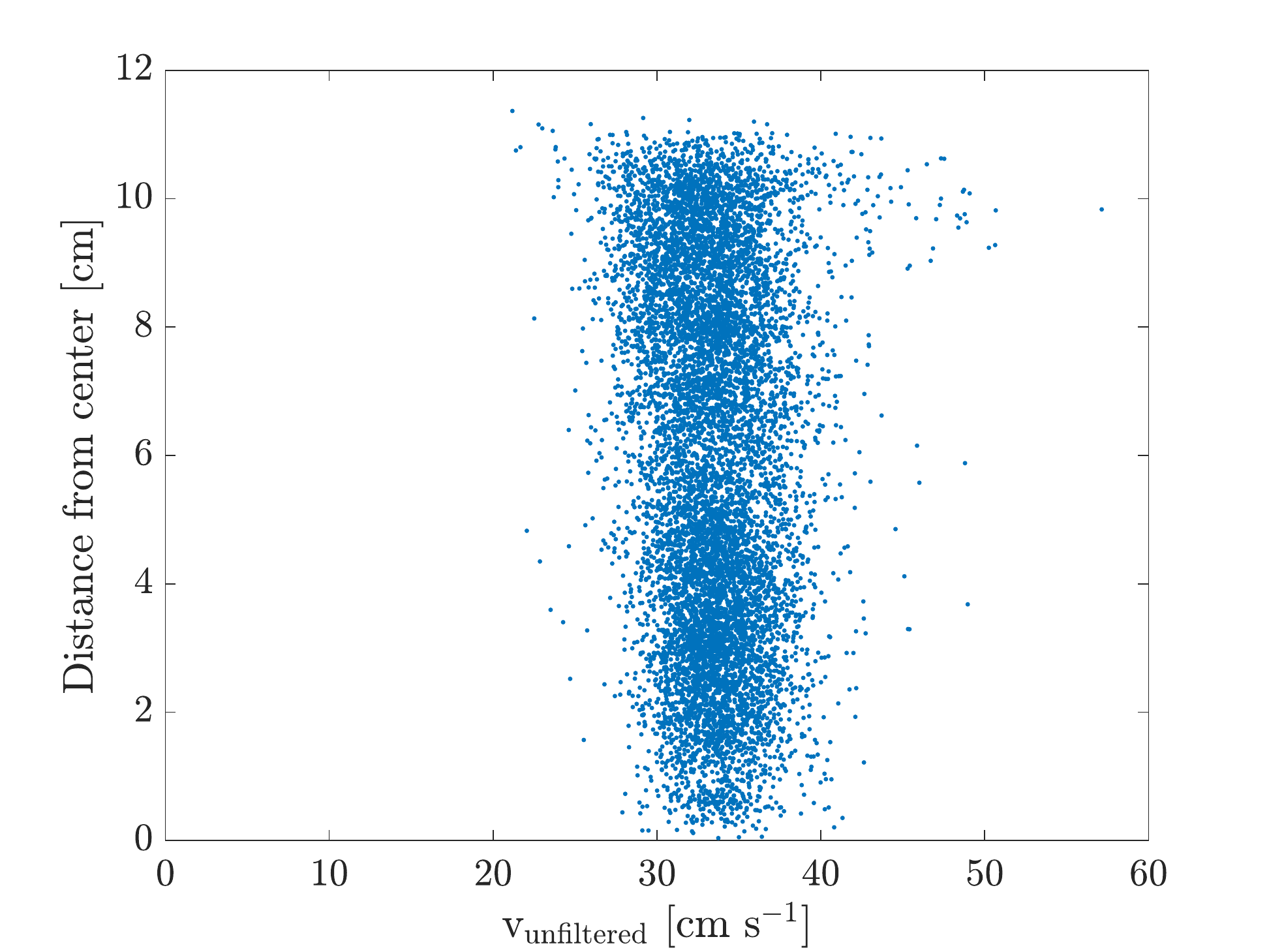}
    \caption{}
    \label{fig:V_afo_R_unfiltered}
  \end{subfigure}
  \begin{subfigure}[b]{0.48\textwidth}
    \includegraphics[width=\textwidth]{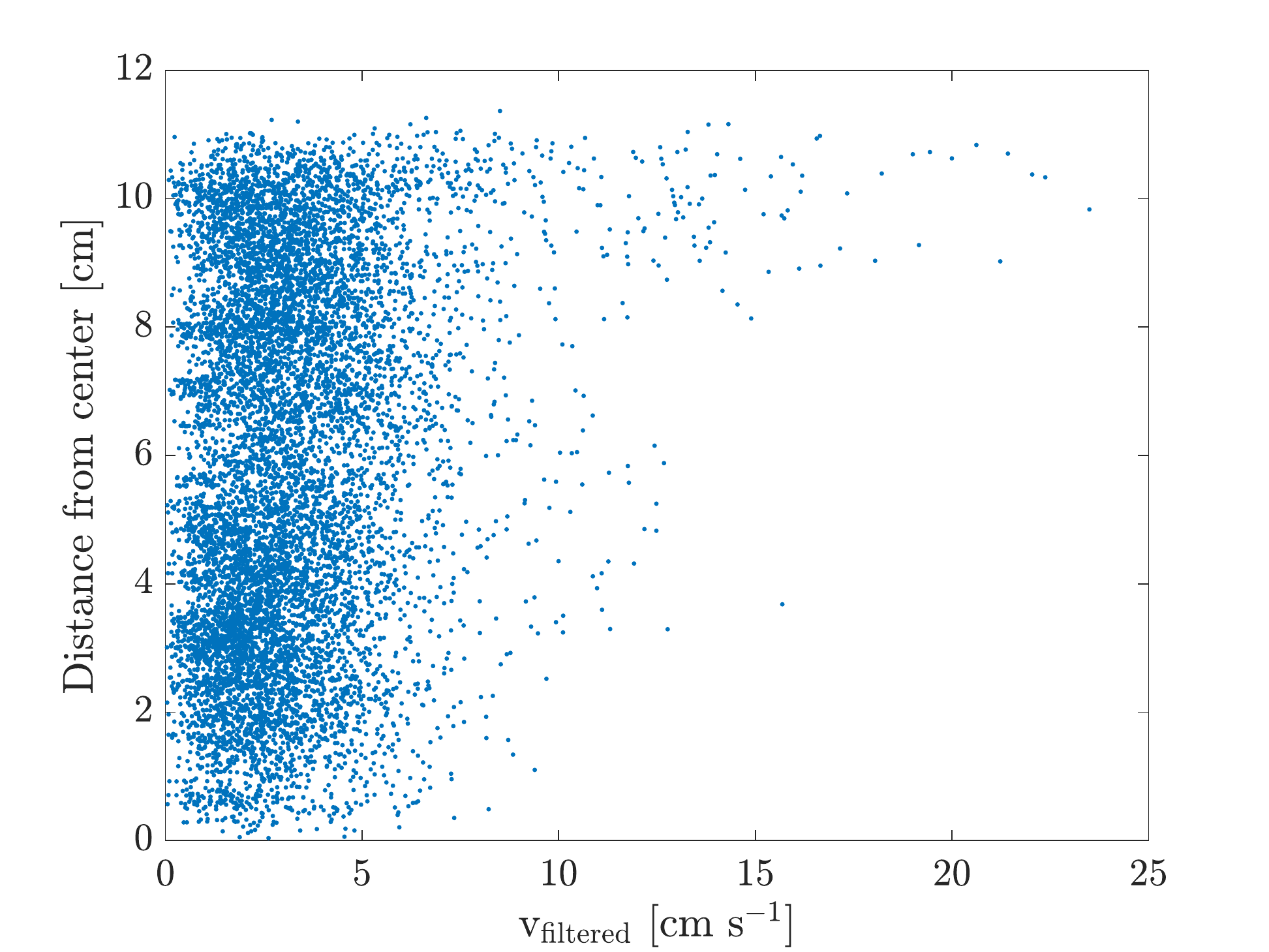}
    \caption{}
    \label{fig:V_afo_R_filtered}
  \end{subfigure}
  \label{fig:V_vs_R}
  \caption{Scatter plots of (a) unfiltered and (b) filtered particle velocity versus particle distance from the centre of the reactor.}
\end{figure*}

\clearpage
\section{Individual particle velocity distributions}\label{app:velocity_distribution_3particles}
\renewcommand\thefigure{C\arabic{figure}}    
\setcounter{figure}{0}  
\begin{figure*}[h]
    \centering
    \begin{subfigure}[b]{\textwidth}
       \includegraphics[width=\textwidth]{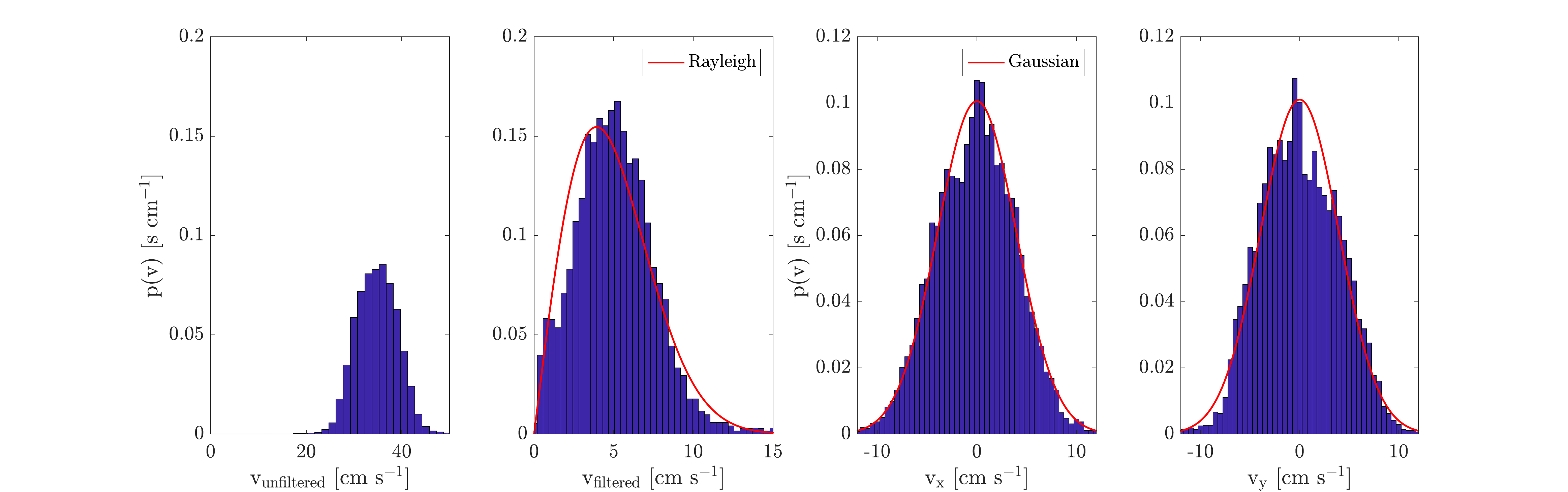}
        \caption{Particle 1}
        \label{fig:particle_1}
    \end{subfigure}
\par\bigskip
    \begin{subfigure}[b]{\textwidth}
       \includegraphics[width=\textwidth]{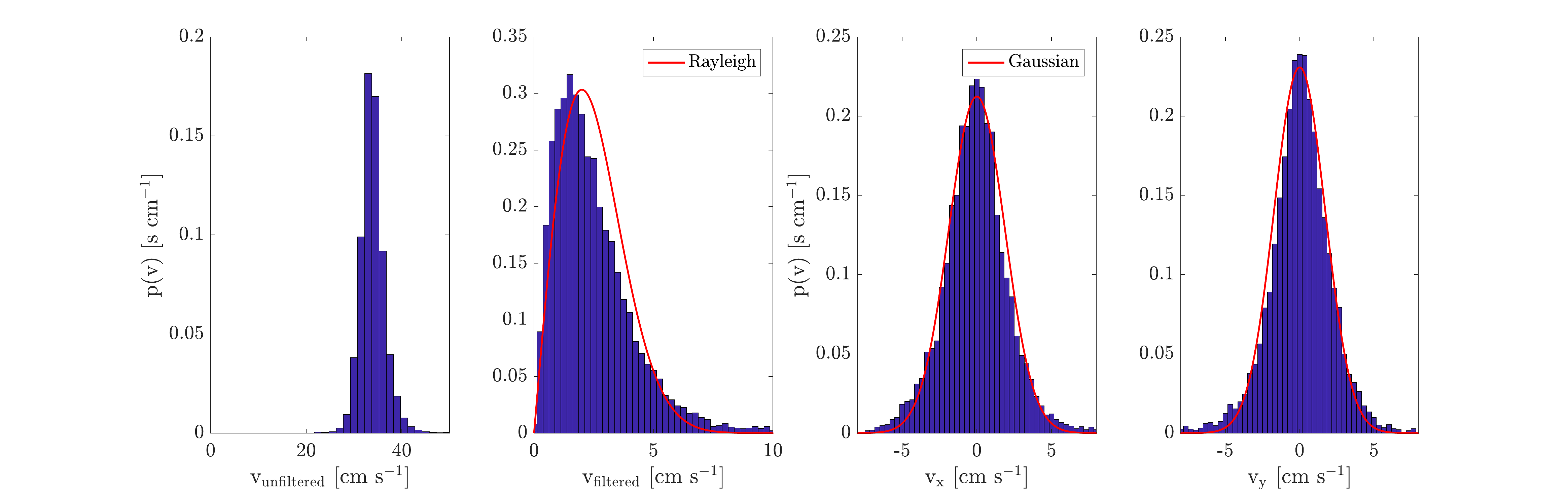}
        \caption{Particle 2}
        \label{fig:particle_2}
    \end{subfigure}
\par\bigskip 
    \begin{subfigure}[b]{\textwidth}
       \includegraphics[width=\textwidth]{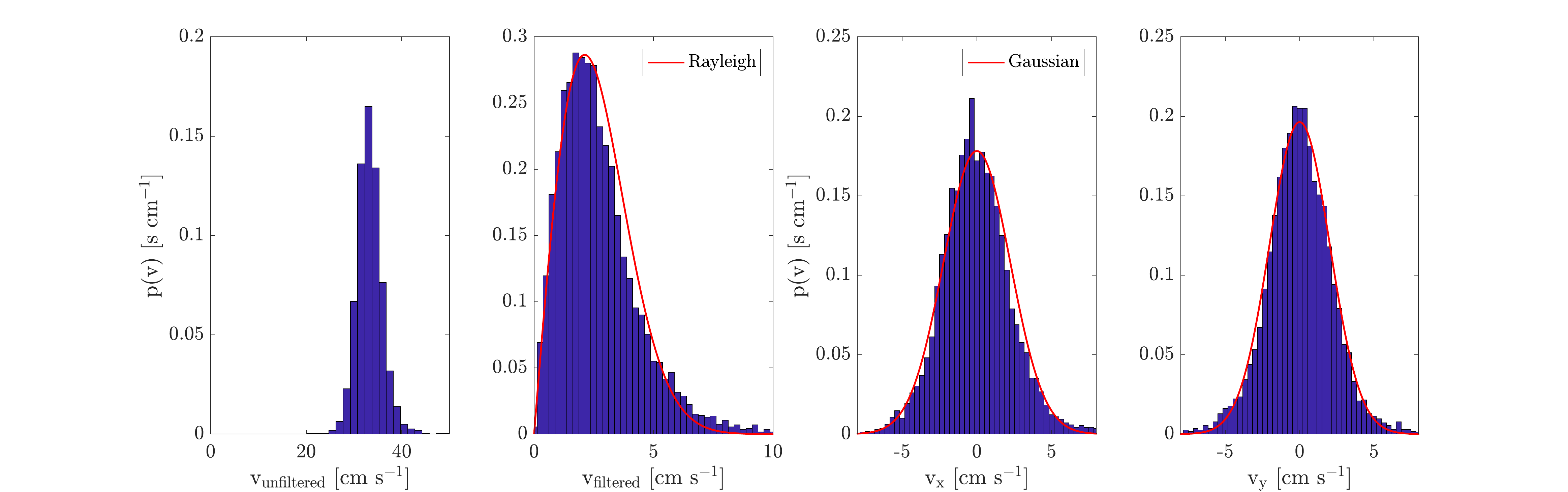}
        \caption{Particle 3}
        \label{fig:particle_3}
    \end{subfigure}
    \caption{Velocity distributions for each particle in the 3-particle experiment. First column: unfiltered 2D velocity distributions; second column: filtered 2D velocity distributions and Rayleigh fitting; third and fourth column: $x$-  and $y$-component 1D velocity distributions and Gaussian fittings. See text for details of the fitting algorithm.}
      \label{fig:velocity_distributions_3particles}
\end{figure*}

\clearpage
\section{Conversion of bounded angle}\label{app:conversion_bounded_angle}
\renewcommand\thefigure{D\arabic{figure}}    
\setcounter{figure}{0}  
\begin{figure}[tbh]
	\centering	\includegraphics[width=0.7\textwidth]{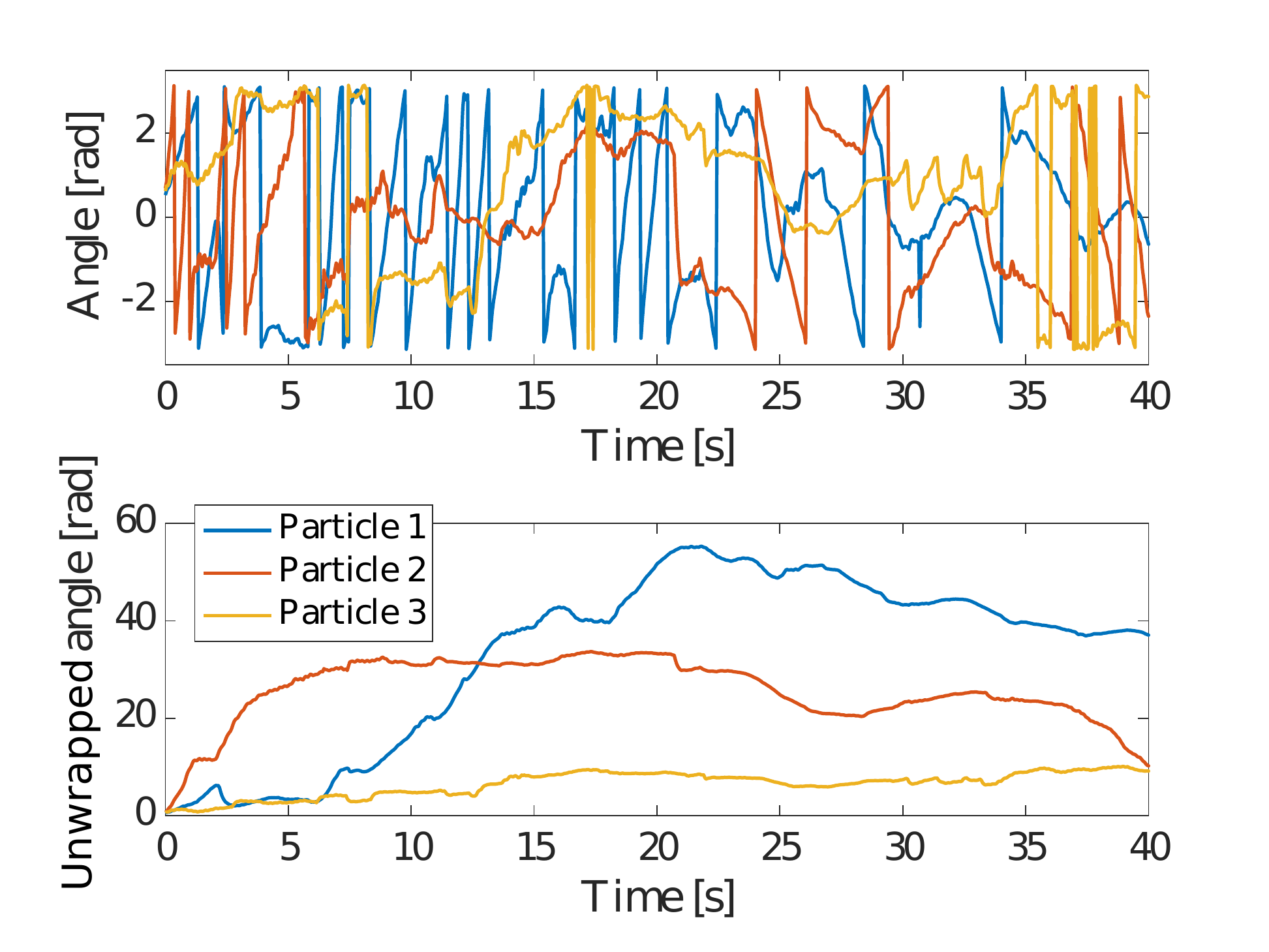} 
	\caption{Conversion of bounded to unbounded angular orientation values for the 3-particle case in the reactor.}
	\label{fig:AngularConversion}
\end{figure}

\end{document}